\documentclass[a4paper,fleqn,usenatbib]{mnras}

\usepackage[T1]{fontenc}
\usepackage{ae,aecompl}

\usepackage[percent]{overpic}

\usepackage{graphicx}	% Including figure files
\usepackage{amsmath}	% Advanced maths commands
\usepackage{amssymb}	% Extra maths symbols
\DeclareMathOperator\erf{erf}

\DeclareMathOperator{\acos}{acos}
\title[Dynamics of cometary irregular particles]{Dynamics 
  of irregularly-shaped cometary particles 
subjected to outflowing gas and solar radiative forces and torques}

\author[F. Moreno et al.]{
Fernando Moreno,$^{1}$\thanks{E-mail: fernando@iaa.es}
Daniel Guirado$^{1}$, Olga Mu\~noz$^{1}$, Vladimir Zakharov$^{2}$,
\newauthor
Stavro Ivanovski$^{3}$, Marco Fulle$^{3}$, Alessandra Rotundi$^{2}$,
Elisa Frattin$^{4}$, Ivano Bertini$^{4}$
\\
$^{1}$Instituto de Astrofisica de Andalucia, CSIC, Glorieta de la Astronomia
s/n, 18008 Granada, Spain\\
$^{2}$INAF - Istituto di Astrofisica e Planetologia Spaziali, Ricerca
Tor Vergata, Via Fosso del Cavaliere 100, 00133 Rome, Italy\\
$^{3}$INAF - Osservatorio Astronomico di Trieste, Via Tiepolo 11,
34143 Trieste, Italy\\
$^{4}$Dipartimento di Astronomia, Universita' di Padova, vicolo
Osservatorio 2, 35122 Padova, Italy
}

\date{Accepted XXX. Received YYY; in original form ZZZ}

\pubyear{2021}

\begin{document}
\label{firstpage}
\pagerange{\pageref{firstpage}--\pageref{lastpage}}
\maketitle

% Abstract of the paper
\begin{abstract}

The dynamics of irregularly-shaped particles subjected to the
combined effect of gas drag and radiative forces and torques in a 
cometary environment is 
investigated. The equations of motion are integrated over distances from the
nucleus surface up to distances where the gas drag is   
negligible.  The aerodynamic forces and torques
are computed assuming 
a spherically symmetric expanding gas. The calculations are limited to 
particle sizes in the geometric optics limit, which is the range of
validity of our radiative torque calculations. The dynamical behaviour
of irregular 
particles is quite different to those exhibited by non-spherical but
symmetric particles such as spheroids. An application of the dynamical
model to comet 67P/Churyumov-Gerasimenko, the target of the {\it
  Rosetta} mission, is made. We found that, for particle sizes larger
than $\sim$10 $\mu$m, the radiative torques are negligible in comparison 
with the gas-driven torques up to a distance of $\sim$100 km from the nucleus
surface. The rotation frequencies of the particles depend on their
size, shape, and the heliocentric distance, while the terminal
velocities, being also dependent on size and heliocentric distance,
show only a very weak dependence on particle shape. The ratio of the
sum of the particles projected areas in the sun-to-comet direction to
that of the sum of the particles projected areas in any direction
perpendicular to it is nearly unity, indicating that the
interpretation of the observed u-shaped scattering phase function by
{\it Rosetta}/OSIRIS on comet 67P coma cannot be linked to mechanical
alignment of the particles. 

\end{abstract}

% Select between one and six entries from the list of approved keywords.
% Don't make up new ones.

%%%%%%%%%%%%%%%%%%%%%%%%%%%%%%%%%%%%%%%%%%%%%%%%%%

%%%%%%%%%%%%%%%%% BODY OF PAPER %%%%%%%%%%%%%%%%%%

\section{Introduction}

In previous papers, \citep{2017Icar..282..333I,2017MNRAS.469S.774I}
have provided a detailed analysis of the dynamics of spheroidal dust
particles in the vicinity of a cometary nucleus by assuming a gas
model characterised by a spherically symmetric expanding
flow. \cite{2017Icar..282..333I} proved that the dynamics of such
aspherical particles is markedly different to spherical particles of
the same volume equivalent radius. The difference in particle shape
and initial orientation on the nucleus surface leads to velocity
dispersion, and the maximum liftable size is minimum for spherical
particles with respect to spheroidal ones. The spheroidal particles,
after some time since ejection from the nucleus, called $t_{rot}$, 
start to experience full rotation, which depends on the particle
physical parameters and nucleus outgassing properties. The model has
been applied to the analysis of {\it{Rosetta}}/GIADA data on comet
67P/Churyumov-Gerasimenko \citep{2017MNRAS.469S.774I}. They found that
the GIADA data are best reproduced with oblate particles rather than
prolate. On the other hand, \cite{2015A&A...583A..14F}, using the
particle tracks as imaged by {\it{Rosetta}}/OSIRIS and a non-spherical
particle model, found that the best agreement between measured and
computed frequencies was reached for oblate spheroids as well.   
 
Based on the argument that the gas drag and the gravitational
attraction of the nucleus 
constitute the dominating forces at distances close to the comet
nucleus, \cite{2017Icar..282..333I} neglected the solar
radiation pressure force and torque in their model. However, 
the effect of the solar radiative force and torque need to be
estimated as they might become 
important at larger ($R\gtrsim5R_N$, where $R_N$ is the nuclear
radius)
distances. We have developed a
 model including the gas 
drag, the gravitational attraction of the nucleus, and the solar
radiation pressure to asses the
long-term evolution of the particles under the combined effect of such
forces. Instead of spheroidal particles, we have considered more
realistic irregularly-shaped particles having a variety of axes
ratios. Irregularly-shaped particles have also been used by
\cite{2014A&A...568A..39C} to 
study meteoroid dynamics. Our
model is based on the rigorous integration of the equation of motion
simultaneously with the Euler dynamical equations, including the
effects of both gas and radiation forces and torques. The
gravitational torques are not included, as they have been shown to be
negligible by \cite{2017Icar..282..333I}.

In Section 2 we provide a detailed description of the equations
involved in the aerodynamic and radiative force and torque models, and
the dynamical equations, where numerical integration is performed
using a quaternion-based scheme. A validation of the computer code is
made through comparison with the results obtained by
\cite{2017Icar..282..333I}, by switching-off the solar radiative force
and torque. In Section 3, results of the full model
including aerodynamic and radiative forces and torques are provided,
as well as a discussion on the combined effect of such forces. We
focus on the evolution of frequencies, degree of tumbling, direction
of the angular momentum vectors, and velocities. We also investigate
the feasibility of the hypothesis of alignment of the largest particle surface
areas respect to the solar radiation, which would lead to backscattering enhancement.  Finally, Section 4 lists the conclusions that can be drawn from the present study.

\section{The Dynamical Model}

This section is divided into three parts. The first part describes the
dust particle models, the second one the outflowing gas model, and the
third part, the radiative force and torque model. As indicated above,
no additional effects, such as the gravitational torque, are taking into account. 

\subsection{Dust Particle Model Build-up}

\begin{figure*}
%\begin{tabular}{ccc}
%	\includegraphics[width=0.33\textwidth]{begonia-eps-converted-to.pdf} &
%        \includegraphics[width=0.33\textwidth]{glasenappia-eps-converted-to.pdf} &
%        \includegraphics[width=0.33\textwidth]{aurora-eps-converted-to.pdf}
  \includegraphics[width=\textwidth]{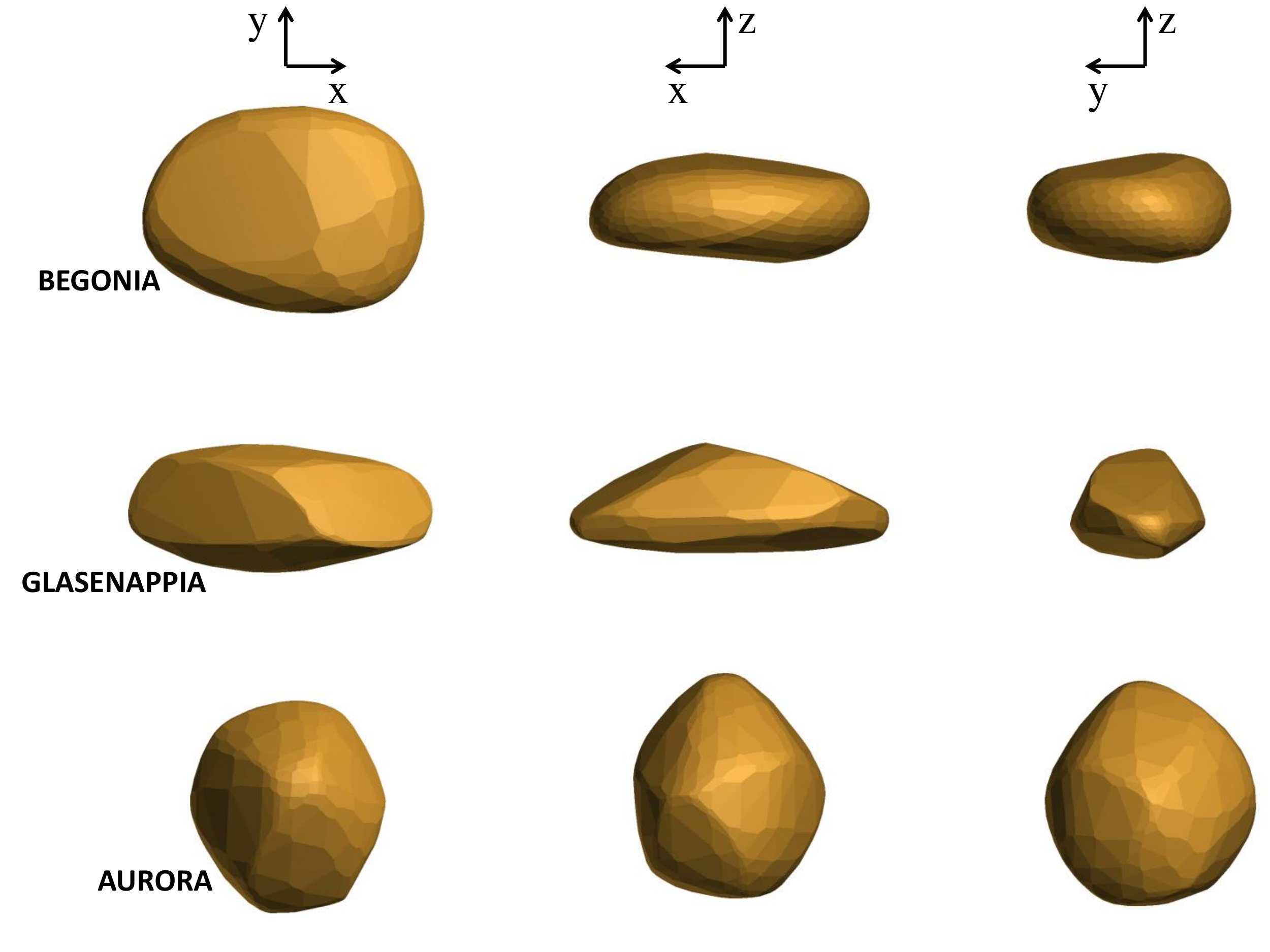}
    \caption{Model particles used in the dynamical
      calculations taken from the 3D Asteroid Catalogue
      (https://3d-asteroids.space/). On the top row of panels, 
      {\it{(943) Begonia}} shape model 
      particle, having a 
      flattened shape, on the middle row {\it{(857) Glasenappia}} shape model
      particle (elongated), and on the bottom row,  {\it{(94) Aurora}}
      shape model particle displaying a rounded shape. Approximate outer
      dimensions of the particles relative 
to $x$-axis are (1.00$\times$0.72$\times$0.36),
(1.00$\times$0.43$\times$0.36), and (1.00$\times$1.09$\times$1.14), respectively.}
    \label{shapes}
\end{figure*}

Irregular shape model particles were adopted from the 3D Asteroid
Catalogue (https://3d-asteroids.space/), which contains 3D models of
known minor bodies derived from lightcurve inversion. We searched the
database to find three distinct kind of shapes, namely flattened, elongated,
and rounded, to explore the differences in their dynamical behaviour. We
selected the shape models of asteroids (943) {\it Begonia}, (857)
{\it Glasenappia}, and (94) {\it Aurora} as representative 
of those different shapes, respectively.  Views of those shape models
displayed with MeshLab \citep{cignoni2008} are given in
figure~\ref{shapes}. In addition, spheroidal 
model particles have also been generated in order to perform the code
validation through comparison with the results by
\cite{2017Icar..282..333I}  (see Subsection 2.4.3). In all cases, the
surfaces are build-up by triangular meshes. 

We assume that the particles behave as solid rigid bodies, they are
homogeneous, and do not contain any volatiles, so that no
rocket forces are applied, and are isothermal, being characterised by
a constant temperature $T_d$. The particle density is set to
$\rho_d$=800 kg m$^{-3}$, which is  
consistent with OSIRIS and GIADA estimates for comet 67P
\citep{2016ApJ...821...19F}. The particle mass is denoted by $m_d$,
related to its effective radius, $r_{e\!f\!f}$ by $m_d =
\frac{4}{3}\pi \rho_d r_{e\!f\!f}^3$. 

\subsection{Aerodynamic force and torque}

For the gas model, we followed the model by
\cite{2017Icar..282..333I}. Briefly, a non-rotating, spherical nucleus, of radius $R_n$
and mass $M_n$ is located at a
certain heliocentric distance $r_h$. The nucleus surface 
temperature is denoted by $T_s$, and is emitting water molecules at a
production rate of $Q_g$. The gas is assumed to behave as an ideal gas
expanding into vacuum, so that it has initial sonic velocity given by
$V_s=\sqrt{\gamma T_s k_B/m_g}$ where $\gamma$ is the ratio of
specific heats, $m_g$ is the
mass of a water molecule, and $k_B$ is the Boltzmann constant. The gas
density ($\rho_g$), velocity ($V_g$), and temperature ($T_g$) are computed as a function of the
distance to the nucleus, $r$, by using the analytical expressions
given by \cite{2018Icar..312..121Z}, appropriate for an adiabatic spherical 
expansion. 
  
The aerodynamic force on the particle is given by \citep[see][]{2017Icar..282..333I}:
\begin{equation}
\mathbf{F_a}=-\int_{S} (p\mathbf{\hat{n}}+\tau\left[(\mathbf{V_r^\prime} \times \mathbf{\hat{n}})
\times
\mathbf{\hat{n}}\right]/|(\mathbf{V_r}\times\mathbf{\hat{n}})\times\mathbf{\hat{n}}
|) \,dS
\end{equation}

where $p$ and $\tau$ are the pressure and the shear stress of the
surface element $dS$, $\mathbf{\hat{n}}$ is an unit vector along the
outer normal to the surface element, and $\mathbf{V_r}$ is the gas to dust
relative velocity.  As already shown by \cite{2017Icar..282..333I},
under the typical physical parameters assumed for the nucleus and the
particles, the flow over the particles can be considered as free
molecular, and the mean collision rate of gas molecules with a dust
particle is always much higher than the
rotation frequency of the particles. The free molecular expressions
for $p$ and $\tau$ can be found in e.g. \cite{1994mgdd.book.....B}, and are given here for completeness as:

\begin{equation}
\begin{split}
p/p_g & = \left[s^\prime \cos \beta/\sqrt{\pi}+\frac{1}{2}\sqrt{\frac{T_d}{T_g}}\right]\exp(-{s^\prime}^2 \cos^2 \beta) \\
&\quad + \left[(\frac{1}{2}+{s^\prime}^2 \cos^2 \beta)+\frac{1}{2}\sqrt{\pi}s^\prime \cos \beta \sqrt{\frac{T_d}{T_g}}\right] \\
&\quad \times \left[1+\erf(s^\prime \cos \beta)\right]
\end{split}
\end{equation}

\begin{equation}
\begin{split}
\tau/p_g & = s^\prime \sin \beta/\sqrt{\pi} \\
&\quad \times \left[ \exp(-{s^\prime}^2 \cos^2 \beta) + \sqrt{\pi} s^\prime \cos \beta \left[ 1+\erf(s^\prime \cos \beta) \right] \right]
\end{split}
\end{equation}

In these equations, $p_g=\rho {{V_r}^\prime}^2/(2{s^\prime}^2)$, $s^\prime={V_r}^\prime\sqrt{m/(2 k_B T_g)}$, and ${V_r}^\prime=V_g - {V_d}^\prime$, where ${V_d}^\prime$ is the velocity of the surface element taking into account the rotation of the particle, and $\beta$ is the angle between  $\mathbf{{V_r}^\prime}$ and $-\mathbf{\hat{n}}$.

Finally, the aerodynamic torque is given by:

\begin{equation}
\mathbf{M_a}=-\int_{S} \mathbf{l} \times (p\mathbf{\hat{n}}+\tau\left[(\mathbf{V_r^\prime} \times \mathbf{\hat{n}})
\times \mathbf{\hat{n}}\right]/|(\mathbf{V_r}\times\mathbf{\hat{n}})\times\mathbf{\hat{n}}
|) \,dS
\end{equation}
where $\mathbf{l}$ is the vector from the particle centre of mass to the surface element $dS$.

\subsection{Radiative force and torque}

\begin{figure}
	\includegraphics[angle=-90,width=\columnwidth]{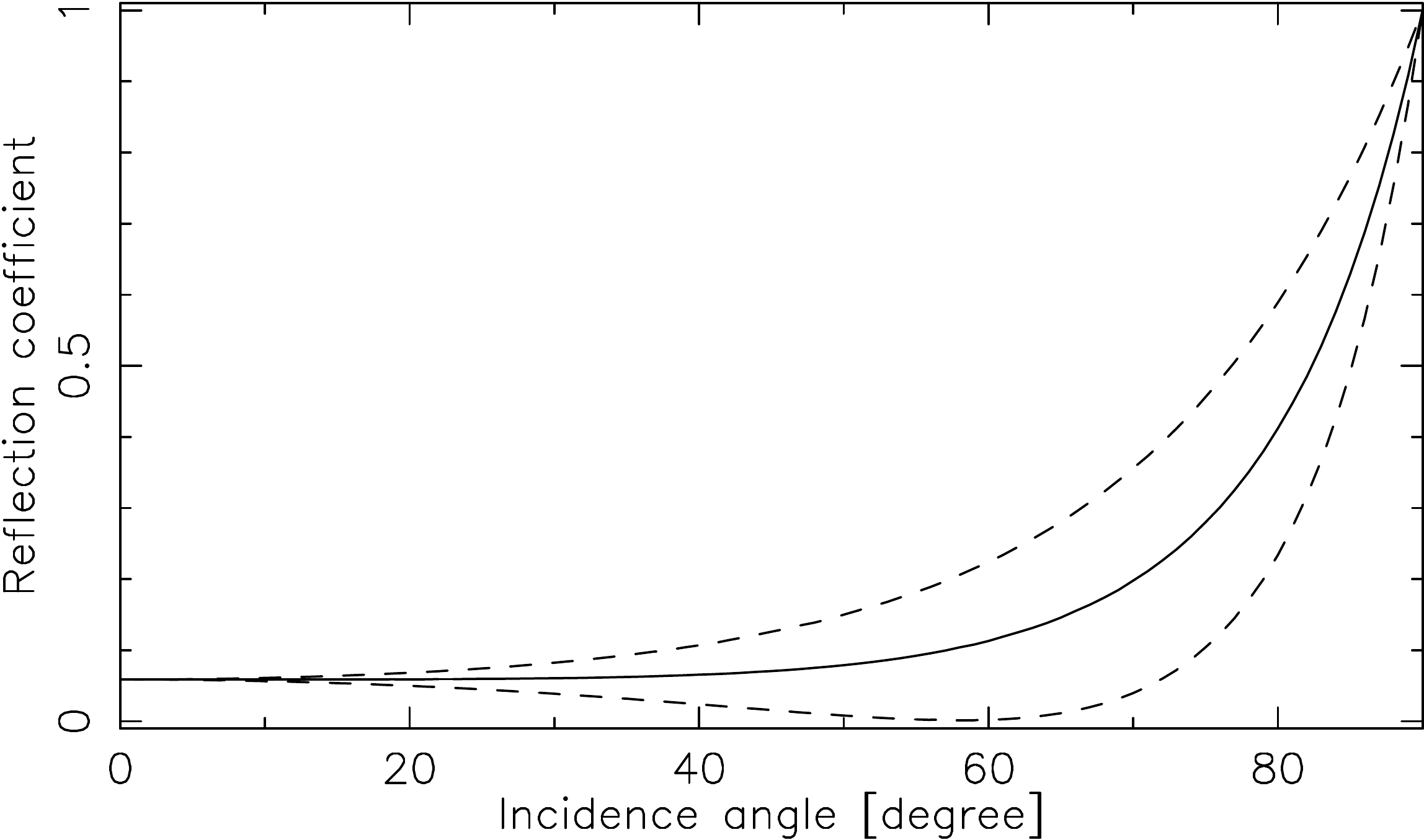}
        \caption{Fresnel reflection coefficients as a
          function of the incidence angle calculated for the
          assumed refractive index $m$=1.6+0.2$i$. The dashed lines
          correspond to the $s$ and $p$ polarizations of the incident
          radiation, and the solid line is the average of those two
          coefficients, which is the value used for $\epsilon_0$ in
          equation (5).}
    \label{ReflectCoeff}
\end{figure}

The calculation of radiative forces and torques on non-spherical particles is a
very CPU and memory consuming task. These quantities can be computed
using light scattering codes such as the Discrete Dipole
Approximation \citep[the DDA code, see][]{1994JOSAA..11.1491D}. The
problem arises when particles of large size 
parameter are considered,
as the DDA method requires a very large number of dipoles
to build up the scatterer (consequently a large memory), and
needs a very 
large CPU time even 
for a single orientation of the particle, so that in practice it
becomes inefficient 
to describe the dynamics of a given particle for large 
integration times. To compute the radiative force and torque on the
particles, knowing that their dimensions are always much larger
than the wavelength of the incident light (assumed at $\lambda$=0.6
$\mu$m), we applied a geometric optics 
approximation. We adopted the expressions
used by \cite{Beletskii66} to compute the force and torque on artificial
satellites by the solar radiation, as follows. The radiation pressure
at a distance $r_h$ from the Sun is given by 
$p_r=\frac{\bar{k}}{r_h^2}$, where $\bar{k}=\frac{E_s}{4\pi c}=1.01
\times 10^{17}$ kg m s$^{-2}$, being $E_s=3.8 \times 10^{26}$ W, the
total power radiated by the Sun, and $c$ the speed of light.  The
solar radiation force on the particle is given by:

\begin{equation}
\mathbf{F_r}=-p_r (1-\epsilon_0) \mathbf{\hat{t}}  \int_{S}
(\mathbf{\hat{t}}\cdot \mathbf{\hat{n}}) dS - 2 \epsilon_0 p_r    \int_{S}
\mathbf{\hat{n}} (\mathbf{\hat{t}} \cdot \mathbf{\hat{n}})^2 dS
\end{equation}

 where $\mathbf{\hat{n}}$ is an unit vector along the outer normal to the
 surface element, $\mathbf{\hat{t}}$ is an unit vector in the opposite
 direction to the solar flux, $dS$ is an elementary surface area on the
 illuminated fraction of the particle, and $\epsilon_0$ is the reflection
 coefficient. The first term in equation (5) corresponds to the
     force exerted by the incident flux and the second term is the
     force produced by the reflected flux. The reflection
 coefficient is computed through  the Fresnel 
 equations for a given complex refractive index $m=n_r+i n_i$. Since
 the incident solar light is unpolarized, the 
 reflection coefficient is calculated as the average of the
 coefficients corresponding to the $s$ and 
 $p$ polarizations (see figure~\ref{ReflectCoeff}).
 
The radiative torque on the particle is given by the following expression:
\begin{equation}
  \mathbf{M_r}= p_r (1-\epsilon_0)\mathbf{\hat{t}}  \times
\int_{S} {\mathbf{l}}(\mathbf{\hat{t}}\cdot \mathbf{\hat{n}}) dS +
2 p_r \epsilon_0 \int_{S} \mathbf{\hat{n}} \times
{\mathbf{l}}(\mathbf{\hat{t}}\cdot \mathbf{\hat{n}})^2 dS 
\end{equation}

where, as before, ${\mathbf{l}}$ is the vector from the particle centre of mass to the surface element $dS$. 
  
To test the validity of this geometric optics approximation, we compared the
torque efficiency resulting from those expressions and those computed
using the DDA code for oblate and prolate spheroids with the largest
possible particle 
dimensions to  
make the geometric optics approximation valid. In all cases, a
refractive index of $m$=1.6+0.2$i$ is used. This value is
    appropriate for cometary particles composed by a mixture of
    silicates and a strongly absorbing component of carbonaceous and
    organic materials, and is in the
    range of values assumed by e.g. \cite{2018ApJ...868L..16M} and
    \cite{2018AJ....156..237M} in their interpretation of the
    Rosetta/OSIRIS phase function measurements of the 67P coma particles.  

The torque efficiency vector is given by \citep{1994JOSAA..11.1491D} as:

\begin{equation}
  \mathbf{Q_M}=\frac{k  \mathbf{M}}{\pi r_{e\!f\!f}^2 u_{r\!a\!d}}
\end{equation}

where $k=2\pi/\lambda$, and $u_{r\!a\!d} =\frac{\bar{k}}{r_h^2}$ is the
energy density due to solar irradiance at the heliocentric distance
$r_h$. The DDA calculations involve the use of dipole arrays obeying
the constraint $|m|kd \le$ 1, where $d$ is the interdipole
distance. This, in practice, limits the particle size to
$r_{e\!f\!f} \lesssim$5$\mu$m for an incident 
wavelength of $\lambda$=0.6 $\mu$m, owing to memory and
CPU available resources. In our tests, we used oblate and prolate 
spheroids having 
 axes ratios of $\epsilon$=0.5 and 2, with $r_{e\!f\!f}$=5$\mu$m,  as
 well as a more extreme 
 case of an oblate spheroid with $\epsilon$=0.125 and
 $r_{e\!f\!f}$=7$\mu$m. The number of dipoles needed to attain $|m|kd \approx$ 
1 varied from 4$\times$10$^6$ to 1.4$\times$10$^7$. The 
calculation involved between 8 and 20 hours of
CPU time per each particle orientation on a Dell\textregistered
workstation with an Intel\textregistered 
Xeon\textregistered E5-1620 3.50GHz processor and 16GB memory. 

\begin{figure}
	\includegraphics[angle=-90,width=\columnwidth]{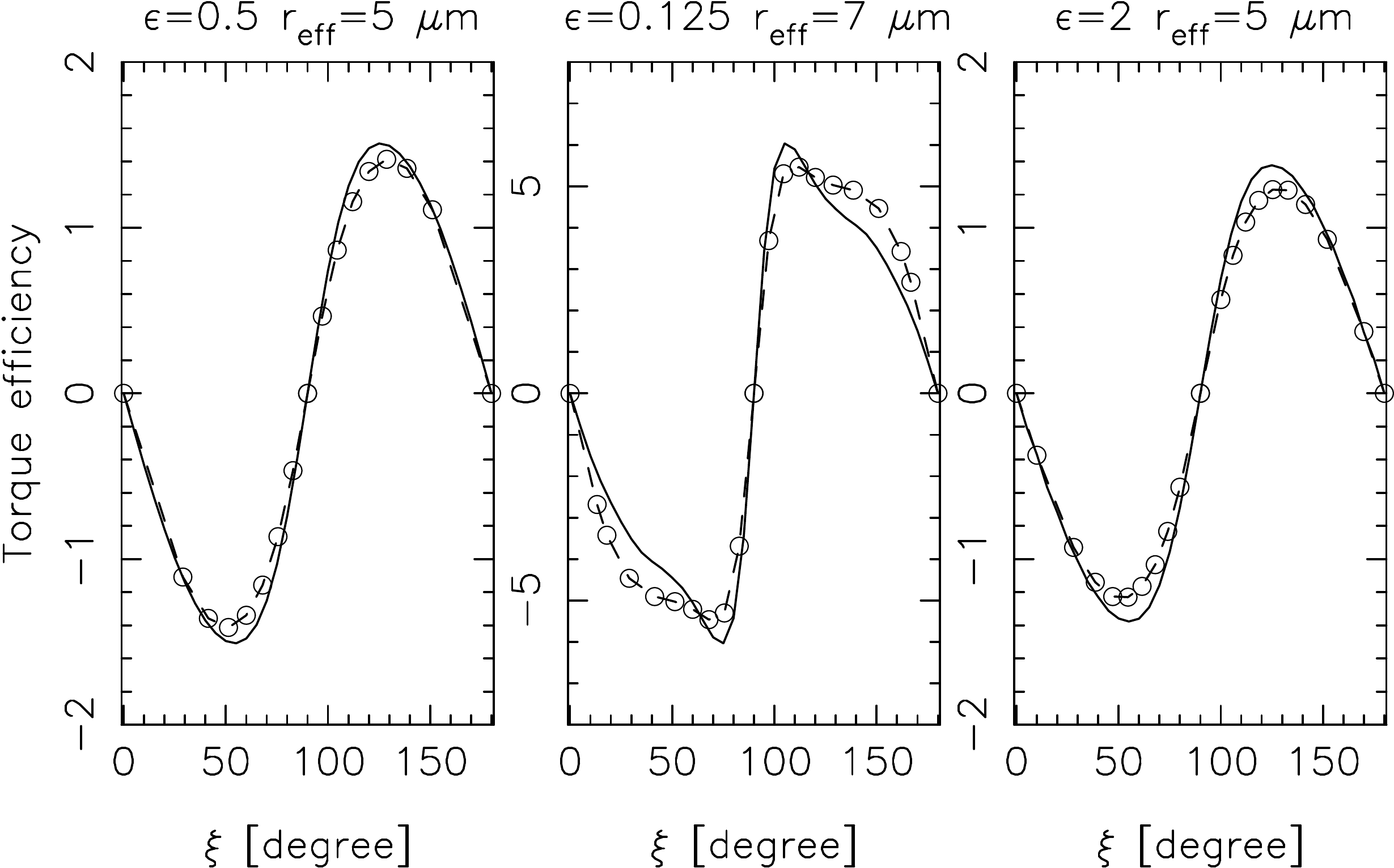}
    \caption{X-component of the torque efficiency as a function of
      the angle of attack $\xi$ (see figure 3) for oblate
      ($\epsilon$=0.5 and $\epsilon$=0.125, left and middle panels) and
      prolate ($\epsilon$=2.0, right panel) spheroidal particles of different
      effective radii, as indicated. The refractive index
      is $m$=1.6+0.2$i$. The solid lines correspond to the geometric
      optics approximation described in Section 2.3, and the open
      circles joined by dashed lines correspond to 
      Discrete Dipole Approximation calculations.}
    \label{radtorques}
\end{figure}

In figure~\ref{radtorques} we display the DDA results together with
our geometric 
optics approximation. The geometry of the problem is depicted in figure 3,
where the direction of the incident flux is along +Z. The X-component of the torque is depicted as a function of the angle of attack
$\xi$. As it can be seen, the agreement between both calculations is
good, even in the case of the oblate spheroid with large aspect
ratio. This allows us to confidently use the geometric optics approximation as a
fast algorithm to compute the radiative torques on the assumed
absorbing and large particles in comparison with the incident
wavelength. It is expected that for still larger particles that those of the
comparison tests, the agreement would be even better, as the
geometric optics regime would be fully attained.

\subsection{Dynamical equations}

We used two reference frames, one centred on the particle centre of
mass, denoted as $(xyz)$, and another centred on the nucleus centre
of mass, denoted by capital letters as $(XYZ)$. The orientation of the
particle with respect to frame $(XYZ)$ is given by the classical Euler
angles, ($\phi$,$\theta$,$\psi$). Rotations of the particle are
defined according to the (3,1,3) (or $zxz$) Euler rotation sequence, also known
as the \textit{x-convention} \citep[see e.g.][]{Diebel06}. 

The translational motion of the particle in the nucleus-centred
reference frame is governed by the equation:
\begin{equation}
m_d\frac{d^2\mathbf{r}}{dt^2} = \mathbf{F_a} + \mathbf{F_r} + \mathbf{F_{gn}} + \mathbf{F_{gs}}
\end{equation}

where we include in the equation the nucleus gravity ($F_{gn}$) and the solar gravity ($F_{gs}$).

The rotation of the particle is described in the $(xyz)$ reference
frame through the Euler equations. The particle axes are set to
    be coincident with its principal 
  axes of inertia, so that the inertia tensor is diagonal
and the Euler equations become: 

 \begin{align}
 I_{xx}\frac{d\omega_x}{dt}+(I_{zz}-I_{yy})\omega_y\omega_z & = M_x \\
 I_{yy}\frac{d\omega_y}{dt}+(I_{xx}-I_{zz})\omega_z\omega_x & = M_y \\
 I_{zz}\frac{d\omega_z}{dt}+(I_{yy}-I_{xx})\omega_x\omega_y & = M_z 
  \end{align}

where $\mathbf{M}=(M_x,M_y,M_z)=\mathbf{M_a}+\mathbf{M_r}$,
$\omega_x,\omega_y,\omega_z$ are the components of the angular
velocity, and $I_{xx},I_{yy},I_{zz},$ are the principal moments of
inertia of the particle. 

To describe the linear and rotational motion of the particle,
equations (8-11) in combination with the Euler kinematic equations
must be numerically integrated.  However, in addition to numerical
instabilities, there are difficulties when directly solving the Euler
equations, such as the well-known Gimbal lock problem \citep[see
  e.g.][]{Zhao13}. Instead, a numerical scheme based on the use of
unit quaternions is much more accurate, and free of Gimbal lock. To
this end, we followed the procedures given by \cite{Zhao13},
which we detailed in the following subsection for
completeness. 

\subsubsection{Quaternions: algebra and relation to rotation matrices}

Quaternions were first introduced by Sir Hamilton (1844), and have been widely applied to solve mechanical problems since the XIX century.  A quaternion $q$ consists of a scalar part and a vector part, $q=\left[q_o, \mathbf{q}\right]$, and is defined as:  
\begin{equation}
q=q_o + q_1 \mathbf{i} + q_2 \mathbf{j} + q_3 \mathbf{k}
\end{equation}
where $q_o, q_1, q_2, q_3$ are real numbers and $(\mathbf{i}, \mathbf{j}, \mathbf{k})$ are unit vectors in the direction of the $(x,y,z)$ axes, respectively.  The conjugate of a quaternion is given by $q^\star =q_o - q_1 \mathbf{i} - q_2 \mathbf{j} - q_3 \mathbf{k}$, while the norm is given by $\|q\|=\sqrt{q_o^2+q_1^2+q_2^2+q_3^2}$. A unit quaternion has norm of unity. The inverse of a quaternion is given by $q^{-1}=q^\star/\|q\|$, so that for a unit quaternion $q$ one has $q^{-1}=q^\star$. 

The multiplication of two quaternions is defined as:
\begin{equation}
t=pq=\left[ p_oq_o  - \mathbf{p}\mathbf{q}, p_o\mathbf{q}+q_o\mathbf{p} + \mathbf{p}\times\mathbf{q}\right]
\end{equation}

In rotation dynamics, a vector $\mathbf{s}$ can be rotated by applying
a rotation matrix, which is a 3$\times$3 matrix $R$ where $R^T=R^{-1}$
and det$(R)$=$\pm$1. The rotated vector  $\mathbf{s^\prime}$, is
written by  $\mathbf{s^\prime}=R\mathbf{s}$. A coordinate (also called
elemental) rotation is a rotation about a given axis. The following
three coordinate rotation matrices rotate a vector by an angle
$\alpha$ about the axis $x$, $y$, $z$:

\begin{equation*}
R_x(\alpha) = 
\begin{bmatrix}
1& 0 & 0 \\
0 & \cos \alpha & -\sin \alpha \\
0 & \sin \alpha & \cos \alpha 
\end{bmatrix}
\end{equation*}

\begin{equation*}
R_y(\alpha) = 
\begin{bmatrix}
\cos \alpha & 0 & \sin \alpha \\
0 & 1 & 0\\
-\sin \alpha &0& \cos \alpha 
\end{bmatrix}
 \end{equation*}

\begin{equation*}
R_z(\alpha) = 
\begin{bmatrix}
\cos \alpha & -\sin \alpha & 0\\
\sin \alpha & \cos \alpha  &  0\\
0 & 0 & 1 
\end{bmatrix}
\end{equation*}
  
Any rotation of a vector $\mathbf{s}$ can be described by a sequence of three coordinate rotations. The most common rotation sequence is the (3,1,3), also known as the x-convention, that is the one that we have used in the modelling. The rotation matrix for a set of Euler angles $(\phi,\theta,\psi)$ in this (3,1,3) sequence is written as $R_{zxz}(\phi,\theta,\psi)=R_z(\phi)R_x(\theta)R_z(\psi)$.

Equivalently, it can be shown that a vector $\mathbf{s}$ can be
rotated by a unit quaternion $q$, resulting in a vector
$\mathbf{s^\prime}$, by the expression \citep[see e.g.][]{Diebel06}:

\begin{equation}
\mathbf{s^\prime} =q\mathbf{s}q^{-1}
\end{equation}

where the multiplication of a vector by a quaternion is simply
    the product of two quaternions where the vector is extended to a
    quaternion having zero scalar part and the same $ijk$ components
    as the vector's $xyz$ components.

Equation (14) implies a relationship between rotation matrices and unit quaternions. In the (3,1,3) sequence, it can be shown that a set of Euler angles $(\phi,\theta,\psi)$ can be converted to a quaternion using the equations \citep{Diebel06}:
 \begin{align*}
q_o &= \cos(\phi/2) \cos(\theta/2) \cos(\psi/2) - \sin(\phi/2) \cos(\theta/2) \sin(\psi/2) \\
q_1 &= \cos(\phi/2) \sin(\theta/2) \cos(\psi/2) + \sin(\phi/2) \sin(\theta/2) \sin(\psi/2) \\
q_2 &= \cos(\phi/2) \sin(\theta/2) \sin(\psi/2) - \sin(\phi/2) \sin(\theta/2) \cos(\psi/2) \\
q_3 &= \cos(\phi/2) \cos(\theta/2) \sin(\psi/2) + \sin(\phi/2) \cos(\theta/2) \cos(\psi/2) 
  \end{align*}
The inverse transformation, quaternion to Euler angles, is given by the equations \citep{Diebel06}:
\begin{align*}
%\phi &= \atan2(2q_1q_3-2q_oq_2,2q_2q_3+2q_oq_1) \\
%\theta &=\acos(q_3^2-q_2^2-q_1^2+q_o^2) \\
%\psi &=\atan2(2q_1q_3+2q_oq_2,-2q_2q_3+2q_oq_1) 
\phi &= \arctan{\frac{2q_1q_3-2q_oq_2}{2q_2q_3+2q_oq_1}} \\
\theta &=\acos(q_3^2-q_2^2-q_1^2+q_o^2) \\
\psi &=\arctan{\frac{2q_1q_3+2q_oq_2}{-2q_2q_3+2q_oq_1}} 
  \end{align*}

\subsubsection{Numerical integration of the dynamical equations}

As stated before, to perform the numerical integration of the dynamical equations, we follow the predictor-corrector method posed by \cite{Zhao13}. 

%VLADIMIR APPROXIMATION:
%
%In assumption that gravity and solar pressure are negligible (probably
%this is quite true for small grains) and $R_N$=const,
%it seems that the terminal translational velocity of dust grains
%$v^t_d$ can be roughly scaled (from grain 0 to grain 1) as:

%\begin{equation}
%\tilde{v}^t_{d1} = \tilde{v}^t_{d0} \sqrt{\frac{Iv_1}{Iv_0}} =
%\tilde{v}^t_{d0} \sqrt{\frac{a_0}{a_1}\frac{Q_{g1}}{Q_{g0}}}
%\end{equation}

%where $\tilde{v}^t_{di}=v^t_{di}/v^{max}_g$, ${\rm Iv}$ is from the
%paper Zakharov et al. Icarus 312 (2018).

%The rotational frequency can be scaled as:

%\begin{equation}
%\tilde{\omega}^t_{d1} = \tilde{\omega}^t_{d0}
%\frac{1-\tilde{v}^t_{d1}}{1-\tilde{v}^t_{d0}}
%\sqrt{\frac{Iv_1}{Iv_0}\frac{a_1}{a_0}} = \tilde{\omega}^t_{d0}
%\frac{1-\tilde{v}^t_{d1}}{1-\tilde{v}^t_{d0}} \sqrt{\frac{Q_1}{Q_0}}
%\end{equation}

%where $\tilde{\omega}^t_{di} = \omega^t_{di} a_i/v^{max}_g$.

%END OF VLADIMIR APPROXIMATION

The particle coordinate axes are set first coincident with the comet (world)
reference frame.  Then, for the irregular particles, at time $t$=0 we
perform an initial random rotation of 
the particle through angles $(\phi_o,\theta_o,\psi_o)$. To perform the
validation tests described below, we used oblate and prolate spheroids
whose initial orientation is simply described by the angle of attack
$\xi$ (see figure~\ref{attack}). The particle is assumed at rest on the comet surface, and having zero angular velocity in the comet reference frame coordinates, i.e., $V_X=V_Y=V_Z=0$ and $\omega_X=\omega_Y=\omega_Z=0$.
For convenience, we assume that the initial coordinates of the
particle are $(0,0,R_n)$. We assume that the sun is placed at a
distance $Z$=$r_h$ from the comet nucleus and located in the plane
$YZ$. The particle is accelerated outwards by the gas drag in the
$+Z$ direction.  In what follows we will denote the vector components
referred to the particle coordinate system with a superscript $\mathbf{b}$
(e.g. $\mathbf{\omega^b}$ or $\mathbf{M^b}$) while no superscript is
used for the variables expressed in the comet, or world reference frame. When indicating the current time level, we used a $\mathbf{n}$ subscript.  The transformation of the torque and the angular velocity from the comet reference frame to the particle reference frame at a given time step $\mathbf{n}$ is given by:

\begin{figure}
	\includegraphics[width=\columnwidth]{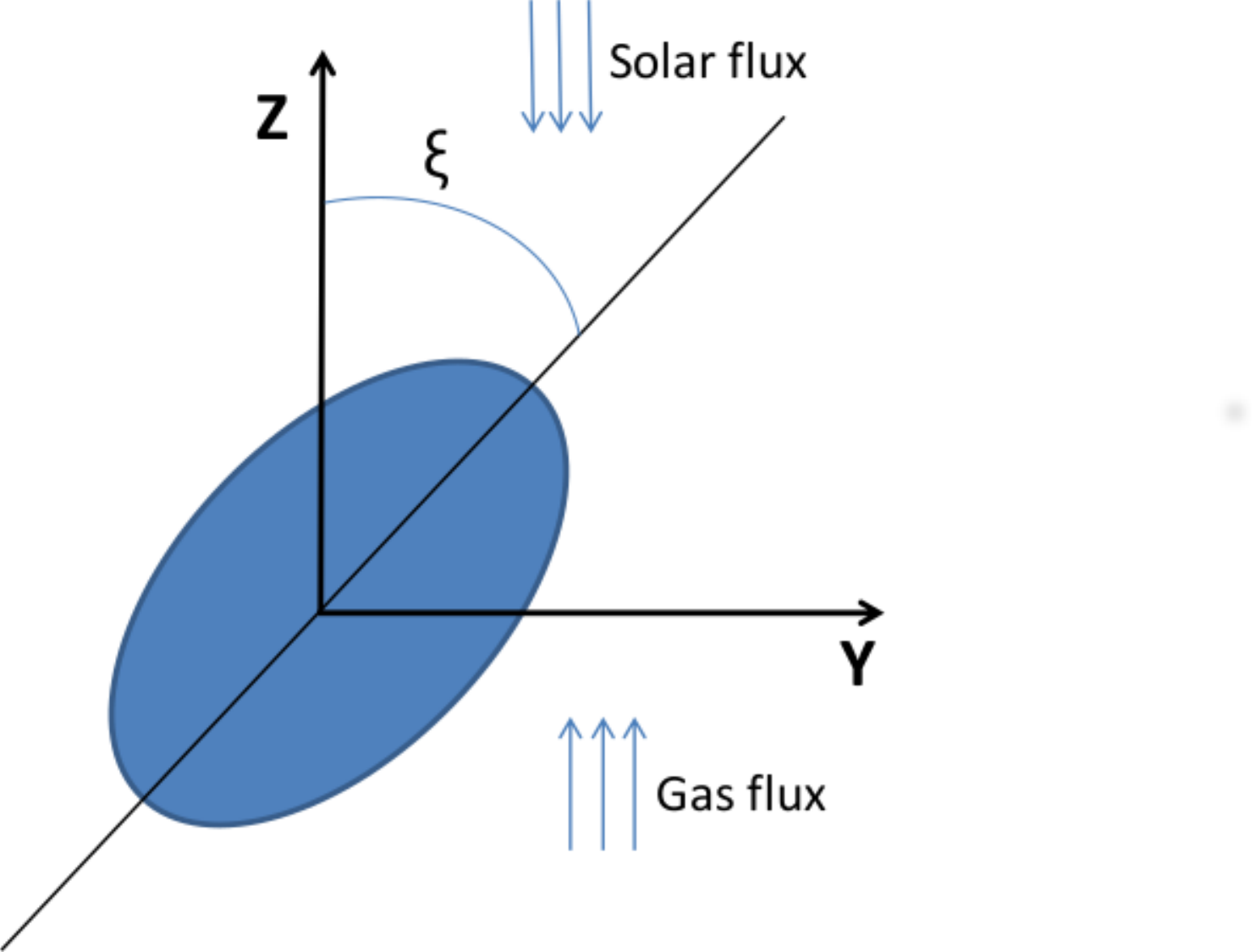}
    \caption{Initial position of a spheroidal particle in the comet
      reference frame. The direction of the gas and solar fluxes
      are along Z-axis. The X-axis points towards the observer
      (right-handed system).  $\xi$ is the angle of attack (see text).}
    \label{attack}
\end{figure}

\begin{align}
 \mathbf{\omega_n^b} &= q_n^{-1}\mathbf{\omega_n}q_n \\
 \mathbf{M_n^b} &= q_n^{-1}\mathbf{M_n}q_n 
 \end{align}
The Euler equations is then used to compute $\dot{\mathbf{\omega}}_n^b$
(equations 8-10). Then, the angular velocities in the particle
reference frame at time $\mathbf{n+\frac{1}{4}}$ and $\mathbf{n+\frac{1}{2}}$ are computed as:
\begin{align}
 \mathbf{\omega_{n+\frac{1}{4}}^b} &= \mathbf{\omega_n^b} + \frac{1}{4}\dot{\mathbf{\omega}}_n^b \Delta t \\
 \mathbf{\omega_{n+\frac{1}{2}}^b} &= \mathbf{\omega_n^b} + \frac{1}{2}\dot{\mathbf{\omega}}_n^b \Delta t 
   \end{align}
The angular velocity in the comet reference frame at time $n+\frac{1}{4}$,  $\mathbf{\omega_{n+\frac{1}{4}}}$, is approximated by the quaternion $q_n$ as:
\begin{equation}
 \mathbf{\omega_{n+\frac{1}{4}}} = q_n \mathbf{\omega_{n+\frac{1}{4}}^b} q_n^{-1}
\end{equation}
A prediction of the unit quaternion at time step $\mathbf{n+\frac{1}{2}}$, $q_{n+\frac{1}{2}}^\prime$ is provided by the following equation:
\begin{equation}
q_{n+\frac{1}{2}}^\prime=\left[\cos \frac{\|\mathbf{\omega_{n+\frac{1}{4}}}\|\Delta t}{4},\sin  \frac{\|\mathbf{\omega_{n+\frac{1}{4}}}\|\Delta t}{4} \frac {\mathbf{\omega_{n+\frac{1}{4}}}}
{ \| \mathbf{\omega_{n+\frac{1}{4}}} \|} \right] q_n
 \end{equation}
Then, the angular velocity in the comet reference frame at time $\mathbf{n+\frac{1}{2}}$
 is obtained as:
 \begin{equation}
   \mathbf{\omega_{n+\frac{1}{2}}} = q_{n+\frac{1}{2}}^\prime
   \mathbf{\omega_{n+\frac{1}{2}}^b} q_{n+\frac{1}{2}}^{\prime-1}
\end{equation}
The torque at time $\mathbf{n+\frac{1}{2}}$ is given by:
\begin{equation}
 \mathbf{M_{n+\frac{1}{2}}} = q_{n+\frac{1}{2}}^\prime \mathbf{M_{n+\frac{1}{2}}^b} q_{n+\frac{1}{2}}^{\prime-1}
\end{equation}
The angular acceleration in the particle reference frame at time $\mathbf{n+\frac{1}{2}}$, $\dot{\mathbf{\omega}}_{n+\frac{1}{2}}^b$ can be then obtained from the Euler equation. 

The corrected quaternion at the next time step $\mathbf{n+1}$ is given by:
\begin{equation}
q_{n+1}=\left[\cos \frac{\|\mathbf{\omega_{n+\frac{1}{2}}}\|\Delta t}{2},\sin  \frac{\|\mathbf{\omega_{n+\frac{1}{2}}}\|\Delta t}{2} \frac {\mathbf{\omega_{n+\frac{1}{2}}}}
{ \| \mathbf{\omega_{n+\frac{1}{2}}} \|} \right] q_n
 \end{equation}
And the angular velocity in the particle reference frame and in the comet
reference frame at time step $\mathbf{n+1}$ can finally be obtained as:
\begin{align}
\mathbf{\omega_{n+1}^b} &= \mathbf{\omega_n^b} + \dot{\mathbf{\omega}}_{n+\frac{1}{2}}^b \Delta t \\
\mathbf{\omega_{n+1}} &= q_{n+1} \mathbf{\omega_{n+1}^b} q_{n+1}^{-1}
\end{align}

At each time step, we simultaneously compute the particle attitude, and
the position and velocity of the particle with respect to the comet
reference frame by equation (8), using the Euler method. 

The computational time step must be much smaller than the inverse of
the rotational frequency of the particle, but much larger than the
gas-grain collisional timescale. We used a time-dependent time step given by 
$\Delta t = 10^{-3} \frac{2\pi}{|\omega|}$ s. As a check
to the validity of the computations, at each time step we perform a
correlation in the three spatial dimensions between the
left- and right-hand sides of the three Euler equations (9-11). We
stop the execution when the correlation coefficient becomes smaller
than 0.99, indicating the presence of instabilities in the
solution. This always occur when the rotational frequencies are too
high so that $\Delta t$ must be set to values even smaller than
those given by the above expression, making the problem intractable in
practice. This problem mostly appears when dealing with small
($r_{e\!f\!f}\lesssim$10 $\mu$m) particle sizes as shown later in Section 3.

\subsubsection{Sample execution of the code: validation and study of the effect of the combined gas plus radiation torques}

\begin{figure}
	\includegraphics[angle=-90,width=\columnwidth]{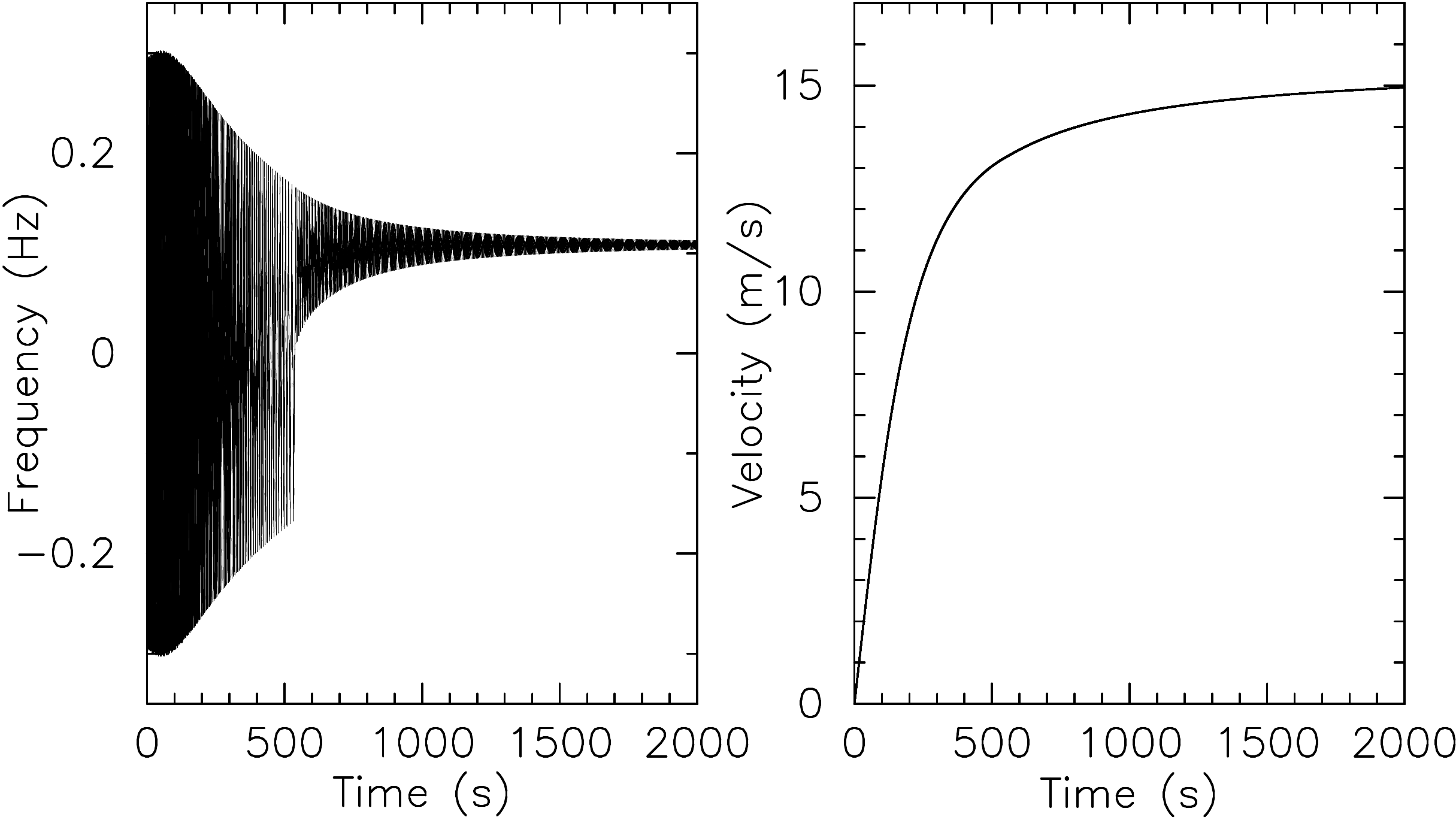}
    \caption{Rotational frequency and velocity as a function of
      time for an oblate 
      spheroid having $\epsilon$=0.5 and effective radius
      $r_{e\!f\!f}$=1 mm. The physical parameters for the comet and
      gas environment are given in table~\ref{tableivanovski}.}
    \label{ivanovski}
\end{figure}

We have validated our code via comparisons with results obtained
by \cite{2017MNRAS.469S.774I}, which refer to gravitational attraction
of the nucleus, and aerodynamic force and
torque on spheroidal particles. For these
comparisons, the model parameters are those of table 1 by Ivanovski et
al., that we reproduce here for completeness in table~\ref{tableivanovski}. The
spheroidal particles experience an oscillatory behaviour first, and
eventually, at time $t_{rot}$, the particle starts to display full
rotation. The asymptotic rotation frequency is denoted by $\nu_{\infty}$. The
particle is accelerated outwards from the comet surface, reaching eventually 
a terminal velocity, $V_{\infty}$. The first
test was made with an oblate spheroid of axis ratio $a/b$=0.5,
effective radius of 1 mm, and density of $\rho$=100 kg m$^{-3}$, 
placed on the comet surface at an angle of attack of 45$^\circ$,
subjected to a water production rate of $Q_g$=10$^{28}$ molecules s$^{-1}$. This
corresponds to case {\it{\#b03g3d1ob}} of table B.9 by Ivanosvki et al. Our
calculations yielded $V_\infty$=15.2 m s$^{-1}$, an asymptotic
frequency of $\nu_{\infty}$=0.107 s$^{-1}$, and a $t_{rot}$=535 s, in excellent
agreement with Ivanovski et al's results ($V_\infty$=15 m s$^{-1}$,
$\nu_{\infty}$=0.1, and $t_{rot}$=534 s). Figure~\ref{ivanovski} displays the
rotational frequency and velocity as a function of time for such 
particle. Additional tests were made for a variety of conditions,
including different production rates, densities, angles of attack, and
shape models (prolate vs oblate). In all cases, the agreement with the
computations of Ivanovski et al. were excellent.

\begin{figure}
	\includegraphics[angle=-90,width=\columnwidth]{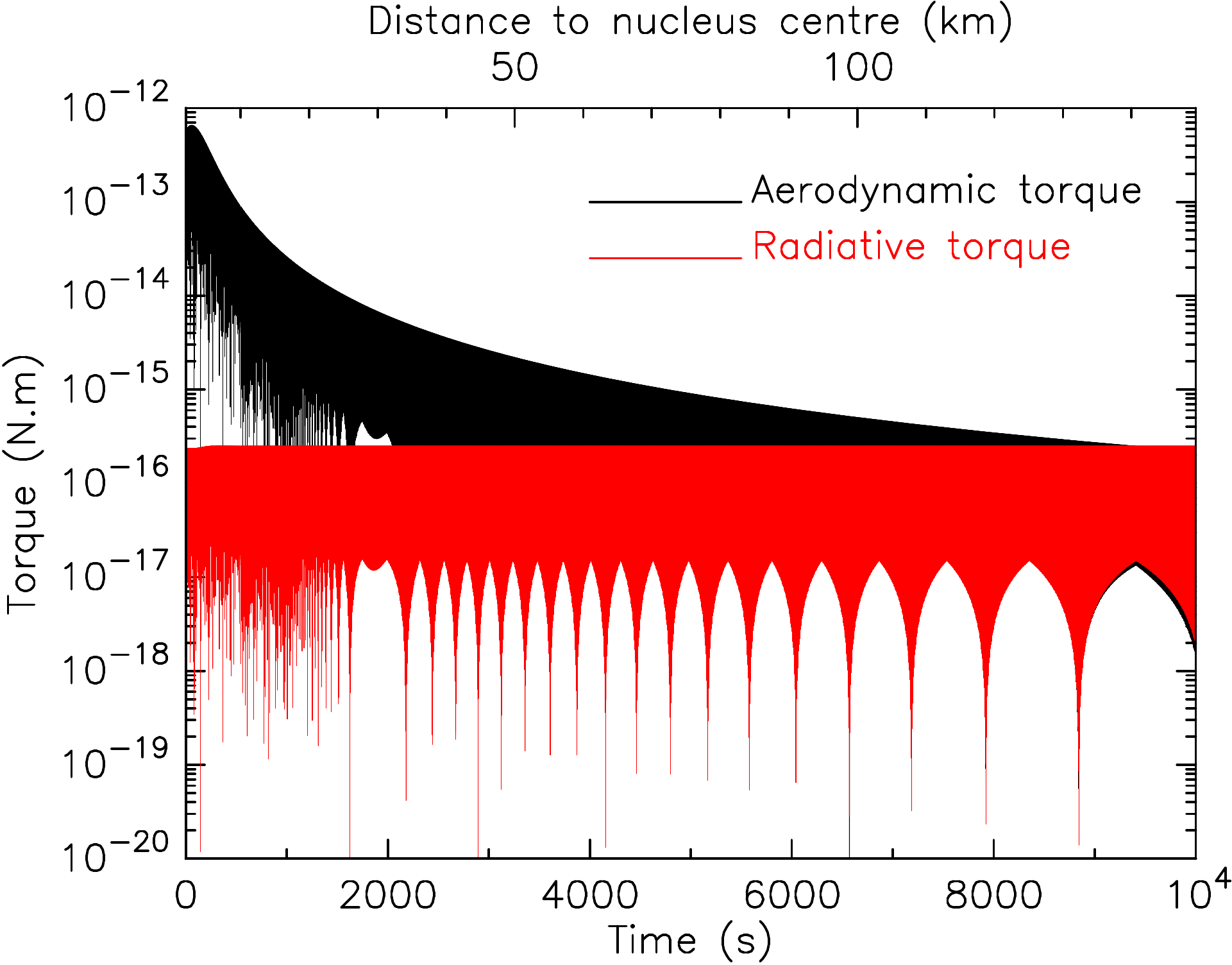}
    \caption{The modules of the aerodynamic (black line) and radiative
      (red line) torques as a function of time and distance to the
      nucleus for the oblate spheroid of $\epsilon$=0.5 and
      $r_{e\!f\!f}$=1 mm described in the text. The physical
      parameters for the comet and 
      gas environment are given in table~\ref{tableivanovski}.}
    \label{torquegasluz}
\end{figure}

\begin{figure}
	\includegraphics[angle=-90,width=\columnwidth]{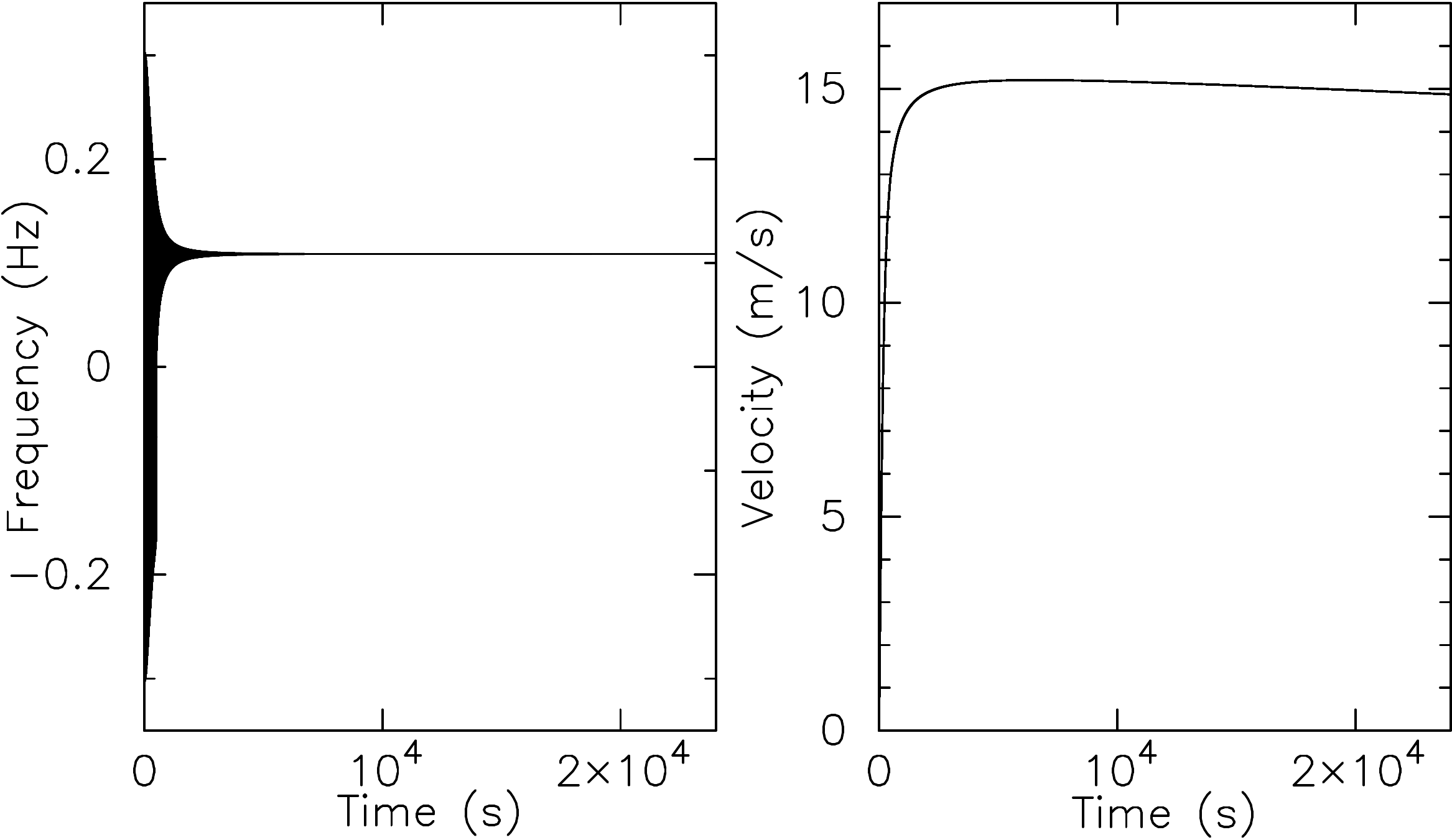}
    \caption{Long-term evolution of the rotational frequency and
      velocity for an oblate  
      spheroid having $\epsilon$=0.5 and effective radius
      $r_{e\!f\!f}$=1 mm. The physical parameters for the comet and
      gas environment are given in table~\ref{tableivanovski}.}
    \label{ivanovski_lux}
\end{figure}

\begin{table}
	\centering
	\caption{Model parameters used for validation with
          Ivanovski et al. (2017b) calculations}
	\label{tableivanovski}
	\begin{tabular}{ll} 
		\hline
		Nucleus radius, $R_N$ [m] & 2000\\
	        Nucleus mass, $M_N$ [kg] & 10$^{13}$\\
	        Nucleus surface temperature, $T_N$ [K] & 200\\
	        Gas composition & H$_2$O\\			
		Gas specific heat ratio, $\gamma$ & 1.33\\
		Particle temperature, T$_d$, [K] & 200\\
		\hline
	\end{tabular}
\end{table}

\begin{figure}
	\includegraphics[angle=-90,width=\columnwidth]{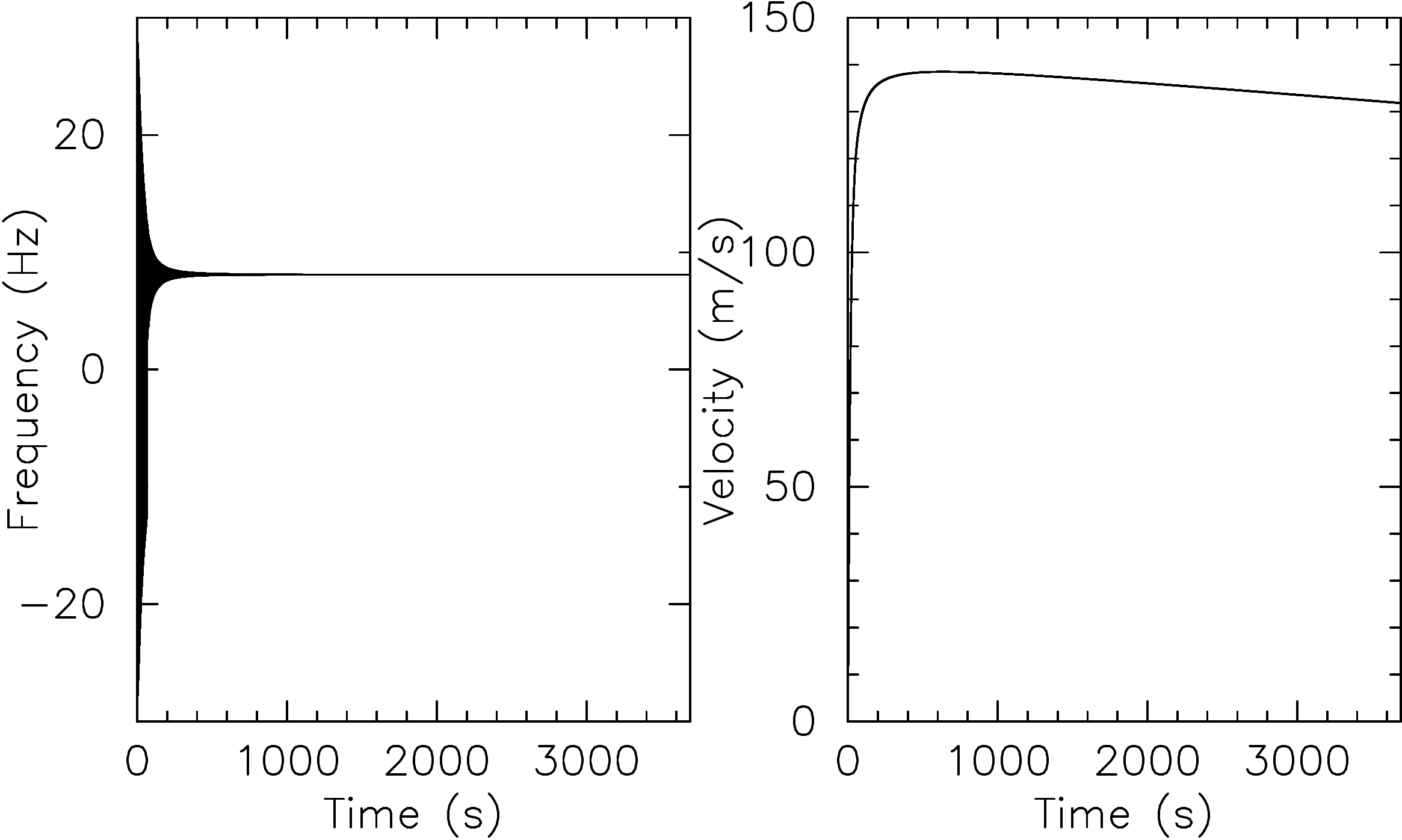}
    \caption{Long-term evolution of the rotational frequency and
      velocity for an oblate  
      spheroid having $\epsilon$=0.5 and effective radius
      $r_{e\!f\!f}$=10 $\mu$m. The physical parameters for the comet and
      gas environment are given in table~\ref{tableivanovski}.} 
    \label{oblate10micron}
\end{figure}

The next step is to show some results aimed at illustrating the
relevance of including the combined effect of the gas drag and
radiative forces and torques. Figure~\ref{torquegasluz} shows the
modules of the aerodynamic and radiative torques on the oblate spheroidal
particle described above. In this graph, we see that the 
aerodynamic torques clearly dominate until  $t\sim$7000 s or to a
distance of the nucleus centre of $R\sim$100 km, where the radiative
torques start to compete with the aerodynamic torques. But even at larger
distances, the effect of the radiative torques on such particle
remains negligible. Figure~\ref{ivanovski_lux} displays the behaviour
of the frequency and particle velocity for an integration time of
2.4$\times$10$^4$ s (i.e., $\sim$360 km from the nucleus), where we see
that the rotation frequency remains unaltered. Only a very slightly
decrease of the velocity, as a consequence of the radiative force,
which is opposite to the gas flux, is noticed. For oblate spheroids of
smaller effective radius of $r_{e\!f\!f}$=10 $\mu$m, the rotation
frequencies encountered are higher, as
expected, but the effect of radiation torques remain negligible, as in
the larger particle size regime. Figure~\ref{oblate10micron} shows the
rotation frequency and velocity for such small particle, were we see
that the asymptotic frequency is $\sim$8 Hz, and the velocity, after
reaching a maximum value near 140 m s$^{-1}$ decrease because of radiation
pressure to $\sim$130 m s$^{-1}$ at the latest position recorded
($\sim$500 km from the nucleus at $t$=3690 s).

We underline that the effect of the
radiative torque on the particles, although being negligible compared
to the expanding gas torques in the nucleus vicinity, they might
become important on longer timescales. However, on long timescales
other processes, such as the Yarkovsky-O'Keefe-Radzievskii-Paddack
(YORP) effect \citep[see e.g.][]{2000Icar..148....2R}, or the Poynting-Robertson drag, become 
important as well. The building-up of a model taking into account
these other forces is well beyond the scope of this work.   

\section{Results of the dynamical modeling}

The purpose of this section is first to show the results of the dynamical
models applied to the three irregularly-shaped particles of
figure{\ref{shapes} under different gas and solar radiation
  environments, and for a variety of effective sizes. Then, in a
  dedicated subsection, we
  build-up a larger data base of particles in a wide
  axis ratio distribution to study their dynamical evolution in order to asses more firmly the
  dynamical parameters. In order to keep 
  the number of free parameters of the model to a 
minimum, we will restrict ourselves to the results relevant to the
physical environment of comet 67P/Churyumov-Gerasimeko, as it has been
subject of much interest because of the successful European Space Agency
mission {\it {Rosetta}}. The comet environmental parameters are those
show in table~\ref{churytable}. The remaining parameters (dust and surface temperatures, $T_d$, $T_S$,
and $\gamma$) are assumed as in
table~\ref{tableivanovski}.

\begin{table}
	\centering
	\caption{Model parameters appropriate for comet 67P/Churyumov-Gerasimenko.}
	\label{churytable}
	\begin{tabular}{lll} 
	  \hline
          Parameter & Value & Reference \\
          \hline
          Nucleus density [kg m$^{-3}$] & 533 &  \cite{2016Natur.530...63P}
          \\
          Nucleus volume [km$^3$] & 18.7 &   \cite{2016Natur.530...63P} \\          
          Nucleus eff. radius [m] & 1650 &  \\
      $Q_g$ [\#s$^{-1}$] ($r_h$=1.24 au) & 3$\times$10$^{28}$
          & \cite{2016MNRAS.462S.491H}\\
      $Q_g$ [\#s$^{-1}$] ($r_h$=2.00 au) & 10$^{27}$ &
          \cite{2016MNRAS.462S.491H}\\          
	         Particle eff. radius [m] &
                 10$^{-5}$ to 10$^{-3}$ &  \\
	         Particle density [kg m$^{-3}$]  & 800 &
                 \cite{2016ApJ...821...19F} \\                 
                 Particle refractive index & m=1.6+0.2$i$ & \\
		\hline
	\end{tabular}
\end{table}

In order to establish the ensemble properties for each particle shape
when subjected to a certain environment, we place the selected particle
shape at the nucleus surface and then perform a random rotation
sequence in the Euler angles $\phi$, $\psi$, and  
$\theta$ according to $\phi=2\pi r_1$,
$\psi=2\pi r_2$, and  $\theta =\acos(1-2r_3)$, where $r_1$, $r_2$, and
$r_3$ are random numbers in the $[0,1]$ interval. We then followed the
trajectory of each particle and record its attitude up to a distance
from the nucleus of 50 km, where the gas drag becomes negligible. The
procedure is then repeated for 100 initial 
positions of each particle shape. However, although our initial purpose was to
accomplish this task for particles having effective radii of 10 $\mu$m,
100 $\mu$m, 1 mm, and 1 cm, the procedure could not be fully completed for the
two smallest sizes (10 $\mu$m and 100 $\mu$m) owing to the presence of
instabilities in
the dynamical evolution that appear for most initial conditions, as
explained below.

For each particle shape and effective radius (1 mm and 1 cm), and for
the two heliocentric 
distances selected, we calculated the degree of
tumbling, {\texttt{DT}}, defined as 
the mean angle between the angular momentum vector and the spin axis
along the integration, and the final direction
of the angular momentum vector $\mathbf{L}_{f\!i\!n\!a\!l}$, the final rotation 
frequency ($\nu_{f\!i\!n\!a\!l}=\sqrt{\nu_x^2 +\nu_y^2 +\nu_z^2}$,
where $\nu=\omega/2\pi$), and the final velocity $V_{f\!i\!n\!a\!l}$ at 
the ending time of integration, when the particles reach a distance
of 50 km from the nucleus centre. The two first parameters, {\texttt{DT}}, and
$\mathbf{L}_{f\!i\!n\!a\!l}$ are computed mainly for the sake of comparison
with the calculations by \cite{2014A&A...568A..39C}.

Tables~\ref{size1mm} and~\ref{size1cm} give the obtained 
parameters for effective radii of 1 mm and 1 cm. In the former case,
the parameters pertain to two heliocentric distances, while in
the latter case, only the parameters at perihelion could be obtained, as
the modelled gas drag is not enough to lift 1 cm particles at $r_h$=2
au.

Concerning the degree of tumbling, we conclude that it depends on the
shape model used, but neither on the size nor on the heliocentric
distance. The rounded shape model, {\it{Aurora}}, shows the smallest
{\texttt{DT}}$\sim$2$^\circ$, while the elongated shape show the largest
{\texttt{DT}}$\sim$25$^\circ$. Interestingly, 
the flattened shape, {\it{Begonia}}, shows the same value 
{\texttt{DT}}$\sim$12$^\circ$ as that 
found by \cite{2014A&A...568A..39C}. Considering all three shape
models together, 
the median of results for $r_{e\!f\!f}$=1 mm at perihelion gives
{\texttt{DT}}=12$\pm$9$^\circ$, while a very similar value for $r_{e\!f\!f}$=1
cm ({\texttt{DT}}=13$\pm$9$^\circ$) is found. At $r_h$=2 au, the median of the
three shape models for $r_{e\!f\!f}$=1 mm gives again a similar result ({\texttt{DT}}=13$\pm$8$^\circ$). 

Another interesting result is the fact that the latitude of the
angular momentum vectors at the end of the integration point to a
latitude interval always centred around
0$^\circ$, i.e., perpendicular to the gas flux
direction. This is also consistent with \cite{2014A&A...568A..39C} findings.

The final frequencies encountered are clearly dependent on the
particle size and shape, and on the heliocentric distance. The largest
frequencies are found for the most elongated shape,
{\it{Glasenappia}}, which vary between 0.2 Hz
for 1 cm particles up to 2.7 Hz for 1 mm particles at
perihelion. For the most rounded shape model, {\it{Aurora}},
notably smaller frequencies are found, moving in the 0.04 to 0.9 Hz
range. The flattened particle rotates at intermediate frequencies
between the other two model shapes. Overall, the values found for
the frequencies of mm to cm sized particles are very consistent with
the those values found 
by \cite{2021MNRAS.504.4687F} in their analysis of a large number of
single particle 
tracks from OSIRIS camera aboard {\it{Rosetta}}. From the lightcurves, 
\cite{2021MNRAS.504.4687F} derived frequencies in the 0.24 to 3.6 Hz
range, with a strong maximum near the lower limit. However, owing to
the limitations of the measurements, the most probable frequency is
likely below the sampled lower limit, which allows us to constrain the 
particle size to be most probably in the cm or higher range. In
addition, many of the particle tracks seen by OSIRIS have associated
lightcurves corresponding to complex rotational motion, as found
here from the irregularly-shaped particles dynamics.   
    
\begin{table}
	\centering
	\caption{Dynamical parameters of $r_{e\!f\!f}$=1
          mm particles at two heliocentric distances.}
	\label{size1mm}
%        \subcaption{Particle properties at perihelion ($r_h$=1.24 au) }
\begin{tabular}{lrrrr}
        \multicolumn{5}{c}{\underline{Particle properties at perihelion ($r_h$=1.24 au)}}\\
	  \hline
\multicolumn{1}{c}{Shape} & \multicolumn{1}{c}{{\texttt{DT}}} & \multicolumn{1}{c}{$\mathbf{L}_{f\!i\!n\!a\!l}$ lat.}  &
\multicolumn{1}{c}{$\nu_{f\!i\!n\!a\!l}$} &
\multicolumn{1}{c}{$V_{f\!i\!n\!a\!l}$} \\
\multicolumn{1}{c}{model} & \multicolumn{1}{c}{(deg)} &
\multicolumn{1}{c}{(deg)} & \multicolumn{1}{c}{(Hz)} &
\multicolumn{1}{c}{(m s$^{-1}$)}   \\
          \hline
              {\it{Begonia}} & 12$\pm$4 & --1$\pm$25 & 1.6$\pm$0.7 & 10.0$\pm$0.2 \\
              {\it{Glasenappia}} & 23$\pm$12 & +4$\pm$32 &
              2.7$\pm$1.1 & 10.0$\pm$0.3 \\
              {\it{Aurora}} & 2.1$\pm$0.7 & +3$\pm$25 & 0.9$\pm$0.5 & 9.34$\pm$0.05 \\
\hline
\end{tabular}
%\bigskip
%\subcaption{Particle properties at $r_h$=2 au}
\begin{tabular}{lrrrr}
        \multicolumn{5}{c}{\underline{Particle properties at $r_h$=2 au}}\\
	  \hline
\multicolumn{1}{c}{Shape} & \multicolumn{1}{c}{{\texttt{DT}}} & \multicolumn{1}{c}{$\mathbf{L}_{f\!i\!n\!a\!l}$ lat.}  &
\multicolumn{1}{c}{$\nu_{f\!i\!n\!a\!l}$} &
\multicolumn{1}{c}{$V_{f\!i\!n\!a\!l}$} \\
\multicolumn{1}{c}{model} & \multicolumn{1}{c}{(deg)} &
\multicolumn{1}{c}{(deg)} & \multicolumn{1}{c}{(Hz)} &
\multicolumn{1}{c}{(m s$^{-1}$)}   \\
\hline
              {\it{Begonia}} & 13$\pm$3  & --5$\pm$23 &
              0.3$\pm$0.2 &
              1.60$\pm$0.05 \\
              {\it{Glasenappia}} & 24$\pm$10 & +5$\pm$32 &
              0.5$\pm$0.3 &
              1.60$\pm$0.05 \\
              {\it{Aurora}} & 2.2$\pm$0.6 & +6$\pm$22 & 0.15$\pm$0.07 & 1.45$\pm$0.01 \\
\hline
\end{tabular}
\end{table}

\begin{table}
	\centering
	\caption{Dynamical parameters of $r_{e\!f\!f}$=1
          cm particles at perihelion ($r_h$=1.24 au).}
	\label{size1cm}
%        \subcaption{Particle properties at perihelion ($r_h$=1.24 au) }
\begin{tabular}{lrrrr}
	  \hline
\multicolumn{1}{c}{Shape} & \multicolumn{1}{c}{{\texttt{DT}}} & \multicolumn{1}{c}{$\mathbf{L}_{f\!i\!n\!a\!l}$ lat.}  &
\multicolumn{1}{c}{$\nu_{f\!i\!n\!a\!l}$} &
\multicolumn{1}{c}{$V_{f\!i\!n\!a\!l}$} \\
\multicolumn{1}{c}{model} & \multicolumn{1}{c}{(deg)} &
\multicolumn{1}{c}{(deg)} & \multicolumn{1}{c}{(Hz)} &
\multicolumn{1}{c}{(m s$^{-1}$)}   \\
          \hline
              {\it{Begonia}} & 12$\pm$4 & 0$\pm$21 &
              0.09$\pm$0.06
              & 3.06$\pm$0.07 \\
              {\it{Glasenappia}} & 26$\pm$10 & +6$\pm$33 & 0.2$\pm$0.1 & 3.10$\pm$0.09 \\
              {\it{Aurora}} & 2.2$\pm$0.5 & +1$\pm$30 &
              0.04$\pm$0.02 &
              2.85$\pm$0.02 \\
\hline
\end{tabular}
\end{table}

Owing to the irregularity of the particles, the rotation frequencies
are always considerably higher than symmetric particles of the same effective
radii and similar aspect ratios. In figure~\ref{example_rotation},
a few typical examples of the evolution of the frequency components and the
rotational energy of particles of different sizes and shapes at the
conditions of 67P perihelion are
shown. In contrast with those symmetric
particles, the rotation is chaotic right after ejection from the
surface, showing a complex rotational motion.  After some distance
from 
the nucleus, the frequency components evolve eventually towards a more regular
pattern, and the rotational energies tend to asymptotic values (see  Figure~\ref{example_rotation}). A similar
behaviour was found by \cite{2014A&A...568A..39C} in his analysis of rotating
meteoroids. \cite{2014A&A...568A..39C} also found a relationship
    between median spin frequencies and velocities for the cases of 
    diffuse and specular reflection of gas molecules on the particle
    surface. For diffuse reflection, he found $\bar\nu \approx 2
    \times 10^{-3} V_{f\!i\!n\!a\!l} D^{-0.88}$, where $D$ is the particle
    diameter. For particles of 1 cm radius and $V_{f\!i\!n\!a\!l} 
    \sim$ 3 m s$^{-1}$ , $\bar\nu \sim$ 0.19 Hz, which is of the same
    order of the frequencies shown in Table ~\ref{size1cm},
    particularly for the {\it{Glasenappia}} shape model particle, and
    also in good agreement with the frequencies obtained for flattened
    and elongated particles having wide axis ratio distribution as 
    described below in Section 3.1 (see Table
    ~\ref{tablelargesample}). For smaller (1 mm radius) particles, the
    predicted frequencies at perihelion and $r_h$=2 au are $\sim$4.5
    Hz and $\sim$0.75 Hz, respectively, which compare well, although
    are bit higher 
    than those shown in Table ~\ref{size1mm}.

As noted above, when the particle effective radius is set down to 100 $\mu$m, in
quite a few cases there appear numerical instabilities that prevent us to
perform any meaningful statistics. The situation becomes worse for
still smaller particles. Successful examples of the 
dynamical evolution of a 
$r_{e\!f\!f}$=100 $\mu$m {\it{Begonia}} shape model particle and of a $r_{e\!f\!f}$=10 $\mu$m {\it{Glasenappia}} shape model particle are given
in Figure~\ref{example_rotation} (lower panels).  For particles having
$r_{e\!f\!f}$=100 $\mu$m, rotation frequencies of order 20-40 Hz can
be reached, while for $\sim$10 $\mu$m particles, frequencies as high
as 500-1000 Hz are observed. In both cases, as for the larger
particles, after a period of time of 
chaotic rotation close to surface, the frequencies evolve to a more
smooth behaviour as the gas drag becomes less and less important. The high
frequencies attained imply that the time-dependent time step
$\Delta t$ had to be as small as $\sim$10$^{-7}$ s in order to keep
the dynamical solution stable during the integration. But even so,  numerical instabilities are seen to appear for most initial
attitudes of the 
particles at a few km from the surface, suggesting that the
integration time step should be  
still smaller, so that the problem becomes intractable in
practice. 

\begin{figure*}
	\includegraphics[angle=-90,width=\textwidth]{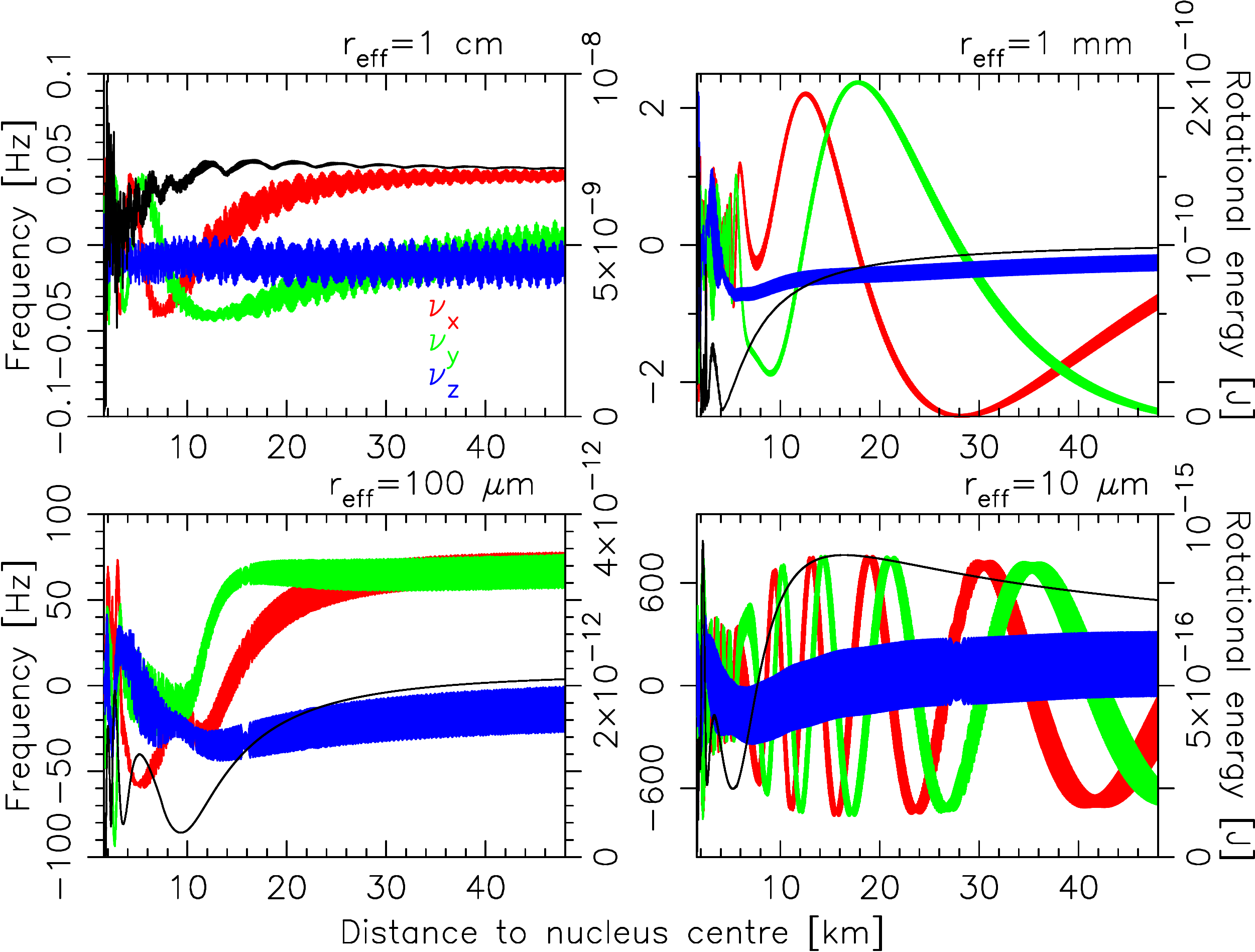}
        \caption{Examples of the evolution of the rotational frequency
          components $\nu_x$, $\nu_y$, and $\nu_z$
      (red, green, and blue lines, respectively, left hand side $y$ axes), and the 
      corresponding rotational energy (black lines, right hand side
      $y$ axes), as a function of 
      distance to the nucleus centre, for various model particles
      with different sizes as indicated. The left panels correspond to
      a {\it{Begonia}}-shaped particle, while the right panels pertain
      to a {\it{Glasenappia}}-shaped particle. The environment
      conditions are those of comet 67P at perihelion (see 
      table~\ref{churytable}).} 
    \label{example_rotation}
\end{figure*}
    
The final velocities, $V_{f\!i\!n\!a\!l}$ are found to depend on
heliocentric distance and size, as expected, but only very slightly on
the particle shape. The maximum values are found for the elongated and
flattened shapes, with values very close between them, and higher
than those shown by the rounded 
shape in the 7-9 per cent interval only. This behaviour is similar
to that shown by spheroidal particles when compared with spheres of
same effective radius \citep{2017Icar..282..333I}. 

As stated above, the numerical simulation for fast rotating particles
requires long 
computational time. On the other side, if we are interested in an
assessment of order of magnitude, it is possible to scale the
available numerical data to other sizes instead of time consuming
numerical simulations. For the scaling we use the relation from
\cite{ivanovski2021} which is based on the approach proposed in
\cite{2018Icar..312..121Z} and further developed in \cite{zvv2021}. This
approach uses a set of universal, dimensionless parameters, which
characterize the dust motion in the inner cometary coma and allows one
to reveal dust flows similarities. 

Table ~\ref{tab_compsc} shows comparison of scaled data with the 
available numerical results for the smallest sizes analyzed, 10
$\mu$m, and 100 $\mu$m. The scaled velocities are in a very good
agreement with the ones calculated numerically. Although the scaled
frequencies differ from the numerical values up to 3-6
times, they still 
provide a reasonable order-of-magnitude estimate, taking into account
the uncertainties in the computed frequencies. The scaled frequencies
for $r_{e\!f\!f\!}$ down to 1 $\mu$m are also given in Table
~\ref{tab_compsc}. This reveals that rotation frequencies in excess of 1000
Hz can be attained for such small particles. A high rotation frequency
could led to particle disruption if centrifugal forces exceed 
the tensile strength. However, following estimates of the critical rotation
period by \cite{1999Icar..142..525D}, the tensile strength of the
material composing the 1 $\mu$m particle rotating at 1000 Hz should be
$\lesssim$10$^{-3}$ Pa to breakup, which is several orders of
magnitude below the current estimates of tensile strengths for
cometary solid particles \citep{2019A&A...630A..24G}.

%%%%%%%%
\begin{table}
\caption{Comparison of scaled and computed values of terminal velocity
$V_{final}$ and frequency $\nu_{final}$.}
\label{tab_compsc}
\centering
\begin{tabular}{l|ll|ll}
\hline\hline
 Shape      & \multicolumn{2}{c|}{$V_{f\!i\!n\!a\!l}$ (m s$^{-1}$)} &
\multicolumn{2}{c}{$\nu_{f\!i\!n\!a\!l}$ [Hz]} \\
            & scaled   & computed & scaled    & computed \\
\hline
\multicolumn{5}{c}{Scaling from $r_{e\!f\!f\!}$=1 mm (Tab.3) to
$r_{e\!f\!f\!}$=100 $\mu$m}\\
\hline
Begonia     & 31.62  &  31.50    & 15.62   &  97.50    \\
\hline

\multicolumn{5}{c}{Scaling from $r_{e\!f\!f\!}$=1 mm (Tab.3) to $r_{e\!f\!f\!}$=10 $\mu$m}\\
\hline
Glasenappia & 100.00  &   98.20   & 243.57  & 780.00    \\
\hline
\multicolumn{5}{c}{Scaling from $r_{e\!f\!f\!}$=1 mm (Tab.3) to $r_{e\!f\!f\!}$=1 $\mu$m }\\
\hline
Begonia     & 316.23 & --       & 1067.13 & --       \\
Glasenappia & 316.23 & --       & 1800.78 & --       \\
Aurora      & 295.36 & --       & 620.24  & --       \\
\hline
\end{tabular}
\end{table}

Another aspect of the dynamics of those rotating particles which is
important to address is whether their associated scattering phase function
would show a strong backscattering enhancement with a minimum near
90$^\circ$ phase angle, based on a purely geometric effect. This
minimum has been clearly found by 
\cite{2017MNRAS.469S.404B} in 67P dust from {\it {Rosetta}}/OSIRIS
images, and the backscattering 
effect has also been seen in several other comets
\citep[see the compilation
  of observations in][]{2017MNRAS.469S.404B}. This phase function
shape has been interpreted with a complex light scattering model 
considering the presence of large particles constituted by 
densely packed submicrometer-sized grains of organic material
and larger micrometer-sized grains of silicate composition
\citep{2018ApJ...868L..16M}. In a previous work
\citep{2018AJ....156..237M}, we provided an alternative interpretation
based on the fact that a combination of large prolate
and oblate shaped 
particles oriented in such a way that their shorter axes point toward
the sun would 
easily explain the 
shape of the observed phase function by OSIRIS as well, without being
strongly dependent on the composition or the structure of the
particles. Laboratory scattering measurements performed by  \cite{2020ApJS..247...19M} with large 
oblate-shaped porous and absorbing particles confirm the theoretical
predictions. With our dynamical model we can perform the calculations
that allows us to check whether this alignment is possible or not. For
this task, we compute  the
average of the ratio
of the sum of the projected areas of a given particle along the sun-comet
line to the sum of the 
projected areas in a perpendicular direction to that sun-comet line,
which we denote as $\Sigma S_0$/$\Sigma S_{90}$.  In
relation to figure~\ref{attack}, $\Sigma S_0$ is computed along $Z$,
and $\Sigma S_{90}$ along any line, selected
randomly, contained in the plane $XY$. Then, if this ratio would
exceed 
unity, this would mean that the particle would be facing a larger 
geometrical cross-section to the sun-comet line (phase angle 0$^\circ$ or 
180$^\circ$), along which most of 
the particles are ejected, than at 90$^\circ$ phase. However, for the three model
particles, we obtained  $\Sigma S_0$/$\Sigma S_{90}$ $\sim$ 1, which
indicates that the phase function backscattering effect observed
cannot be attributed to a mechanical alignment of particles in the coma.   

The dynamical properties derived so far correspond from the model particles
having three single-shaped particles. The question then arises
as to whether the dynamical properties derived are similar or not for 
more realistic wider axes ratio distributions of particles. Thus, in
the next subsection 
we describe computations of the particle dynamics for a large sample of
such particles in order to derive more reliable statistical dynamical parameters.  

\subsection{Dynamics of wide axes ratio distribution of particles} 

Since both the degree of
tumbling and the direction of the angular momentum vector $\mathbf{L}$
seem rather insensitive to particle size and heliocentric distance,  for this analysis we
restricted ourselves to particles of $r_{e\!f\!f}$=1 cm at
perihelion. The study consists 
in generating a large set of particles, placed, as before, at initial
random attitudes on the comet surface to study their dynamical
evolution. We divided the shape models into two families, one
oblate-like, and another prolate-like particles.  To build each model
particle, we first generated 
a cloud of points at random positions on the surface of an ellipsoid
defined by axes $(a,b,c)$. Those axes were  
defined as $(a,b,c)=(6+3r_1,6+3r_2,1+2r_3)$ for the oblate-shaped
particles, and  $(a,b,c)=(1+2r_4,1+2r_5,4+3r_6)$ for the
prolate-shaped ones, where $r_i$ ($i$=1 to 6) are random numbers in the
$[0,1]$ interval. This would yield $c/a$ between 0.1 and 0.5 for the
oblate-shaped and $c/a$ between 1.3 and 7.1 for the prolate-shaped.
Then, from each cloud of random points we
generated a convex hull surface, and imposed a quadric
edge collapse decimation to simplify the surface to 40 faces
each, using MeshLab \citep{cignoni2008}. We computed subsequently the 
inertia tensor of the shape model generated, that was diagonalised by
Jacobi's method. The integration of each particle was performed up to
a distance of 50 km from the nucleus centre, as with the
asteroidal-shaped particles. Examples of the shape models generated are 
shown in Figure~\ref{examplesynth}. The total number of model particles
were 3500 for those oblate-shaped and 2200 for the
prolate-shaped.      

\begin{figure}
\centering
\begin{tabular}{cc}
	\includegraphics[width=0.5\columnwidth]{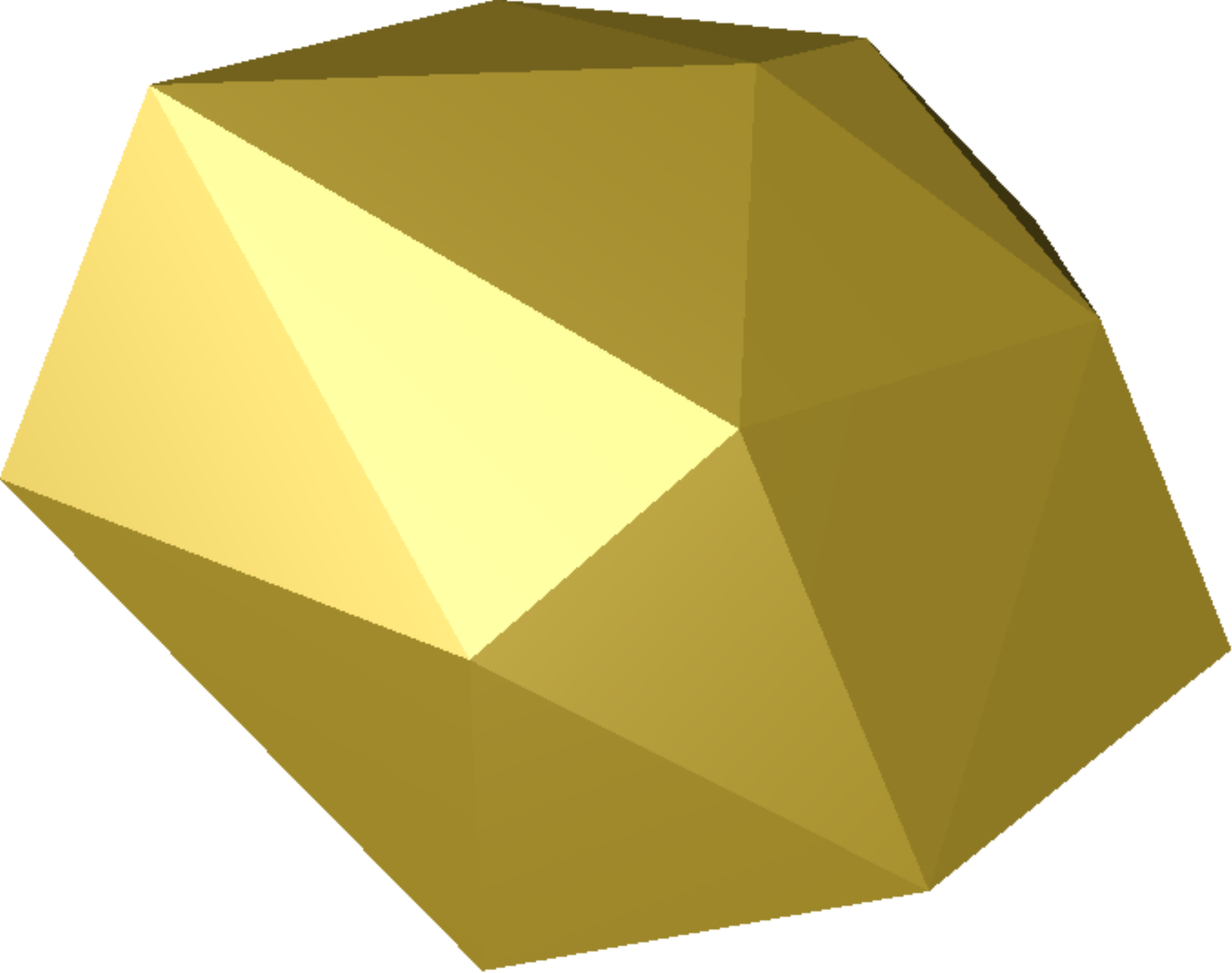} &
        \includegraphics[width=0.5\columnwidth]{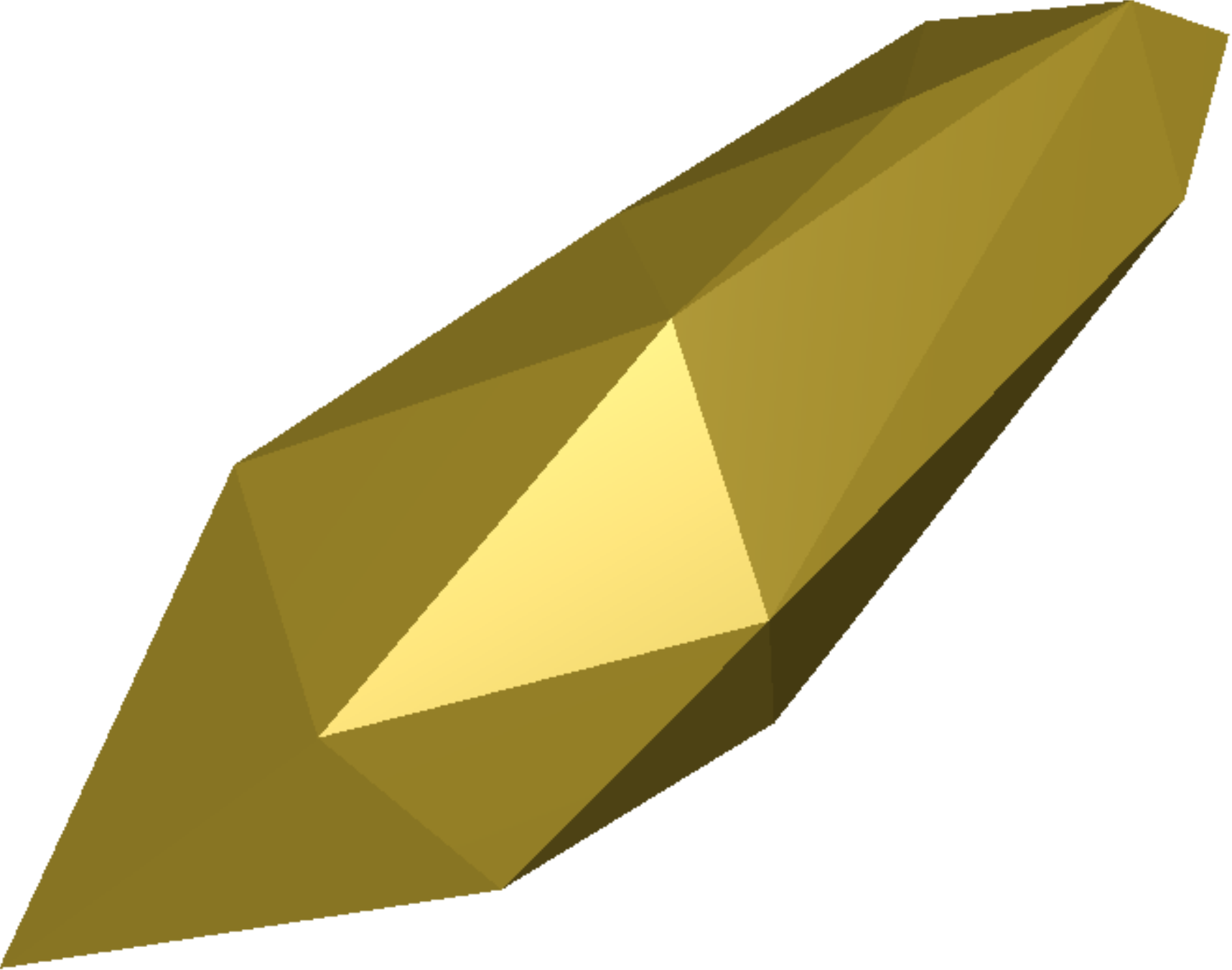}  
 \end{tabular}
    \caption{Flattened (left) and elongated (right) synthetic 
      sample model particles.}
    \label{examplesynth}
\end{figure}

%\begin{figure*}
%\centering
%\begin{tabular}{ccc}
%	\includegraphics[angle=-90,width=0.33\textwidth]{ldir_ob-eps-converted-to.pdf} &
%        \includegraphics[angle=-90,width=0.33\textwidth]{tumbling_ob-eps-converted-to.pdf} &
%        \includegraphics[angle=-90,width=0.33\textwidth]{sumas_ob-eps-converted-to.pdf} \\
%	\includegraphics[angle=-90,width=0.33\textwidth]{ldir_pr-eps-converted-to.pdf} &
%        \includegraphics[angle=-90,width=0.33\textwidth]{tumbling_pr-eps-converted-to.pdf} &
%        \includegraphics[angle=-90,width=0.33\textwidth]{sumas_pr-eps-converted-to.pdf} \\
% \end{tabular}
%    \caption{Flattened (left) and elongated (right) synthetic shape
%      model particles.}
%    \label{largesample}
%\end{figure*}

\begin{figure*}
\centering
\begin{tabular}{ccc}
\begin{overpic}[angle=-90,width=0.33\textwidth]{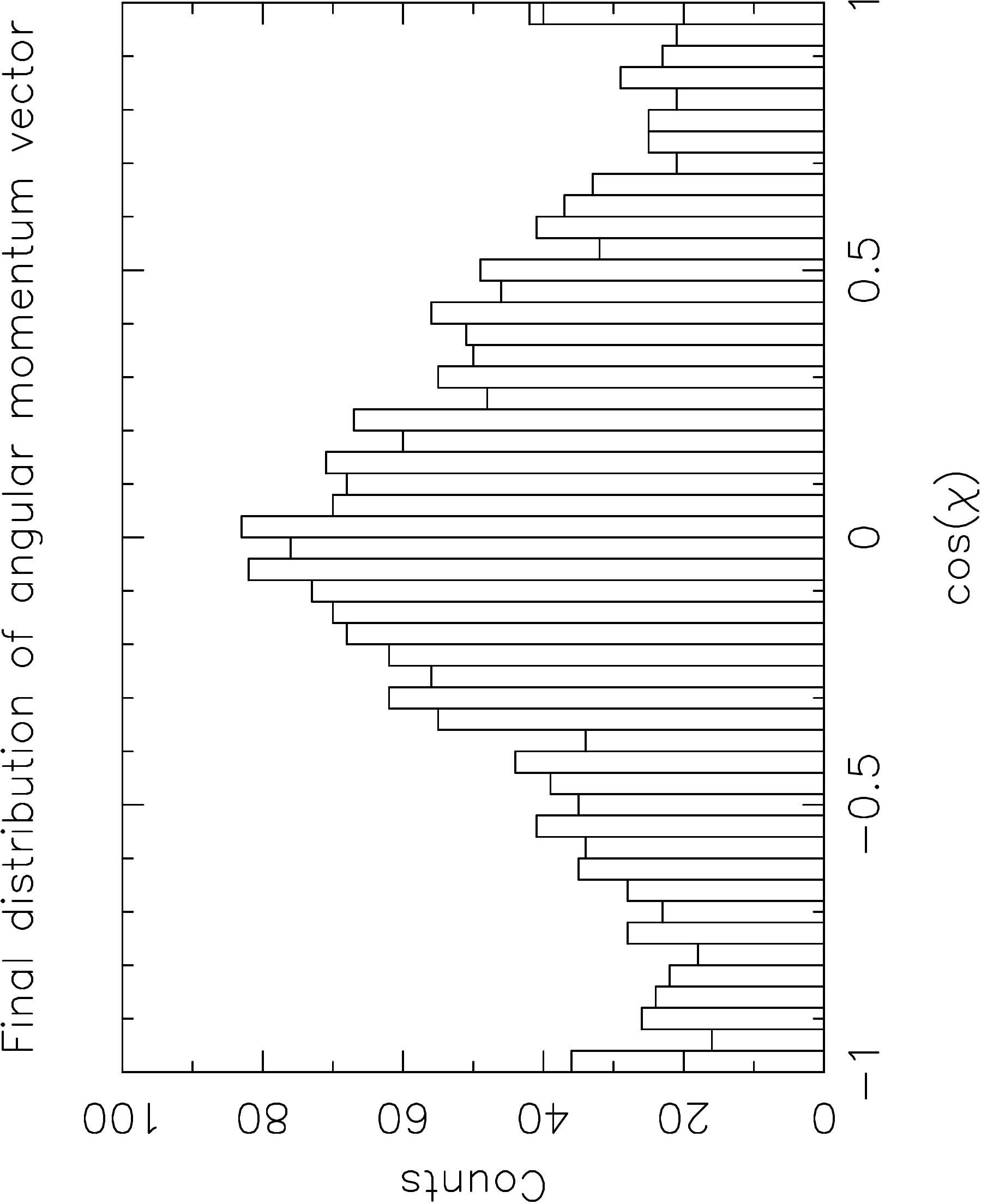}
\put (67,62) {\large OBLATE}
\end{overpic} &
\begin{overpic}[angle=-90,width=0.33\textwidth]{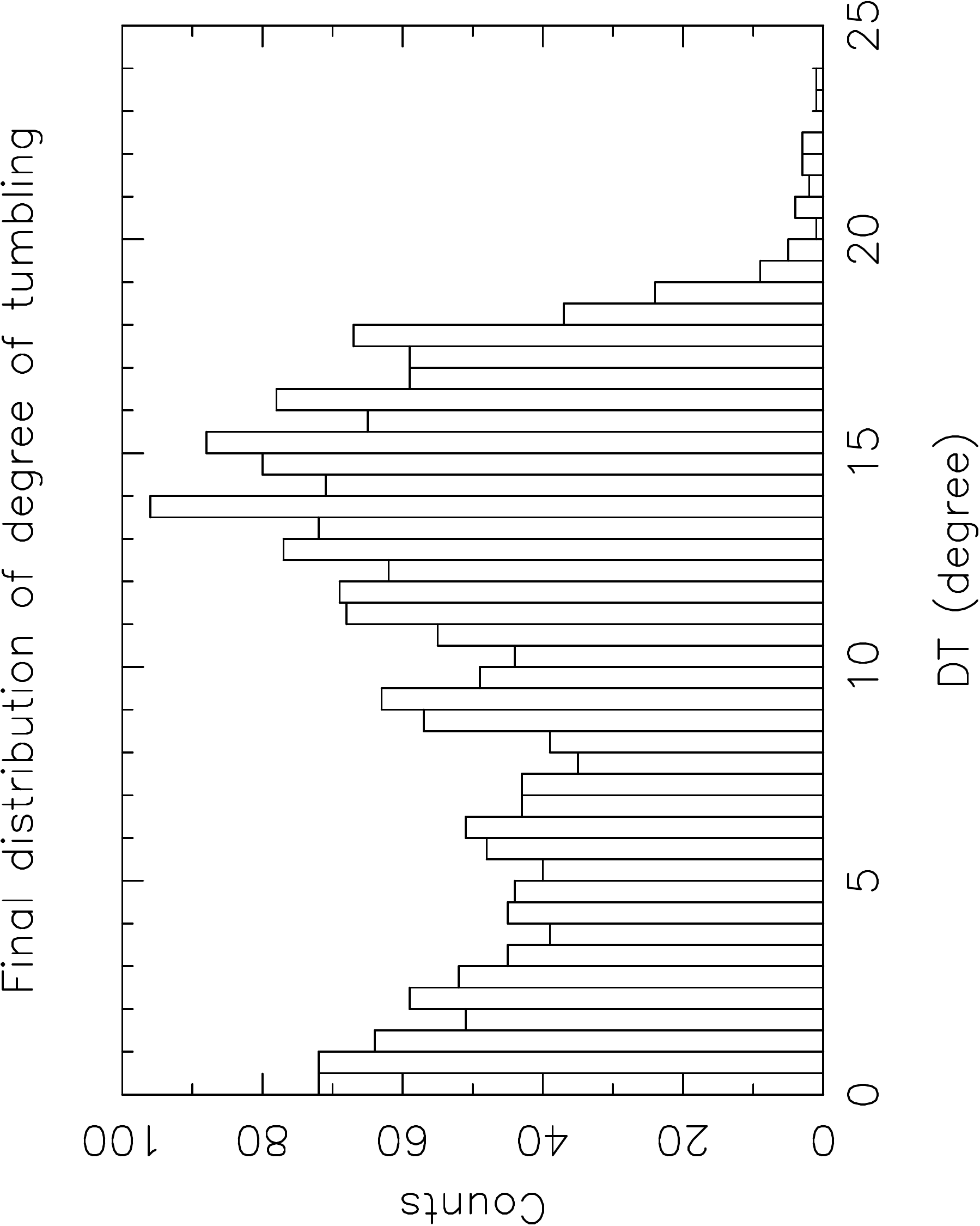}
\put (67,62) {\large OBLATE}
\end{overpic} &
\begin{overpic}[angle=-90,width=0.33\textwidth]{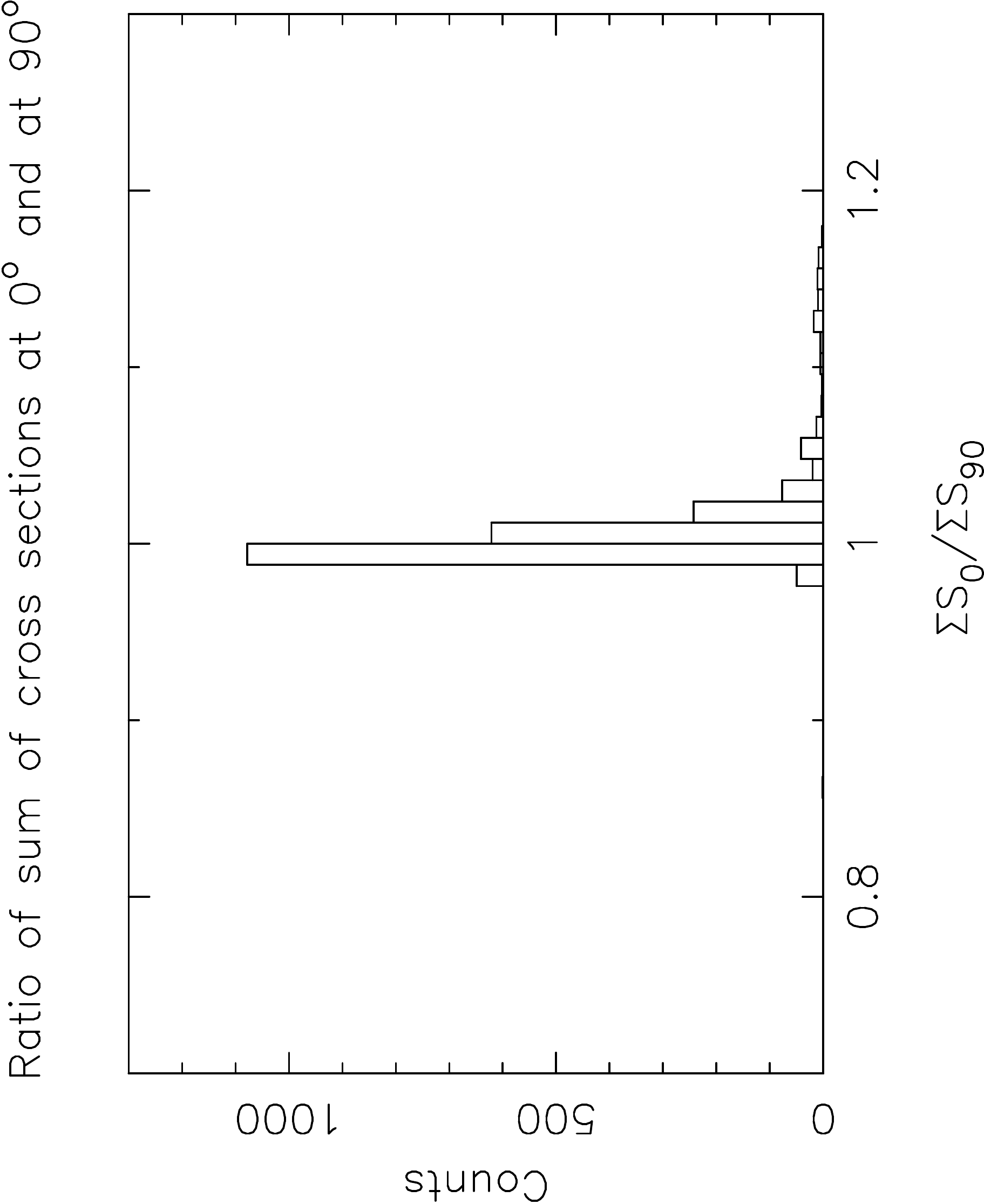}
\put (67,62) {\large OBLATE}
\end{overpic} \\
\begin{overpic}[angle=-90,width=0.33\textwidth]{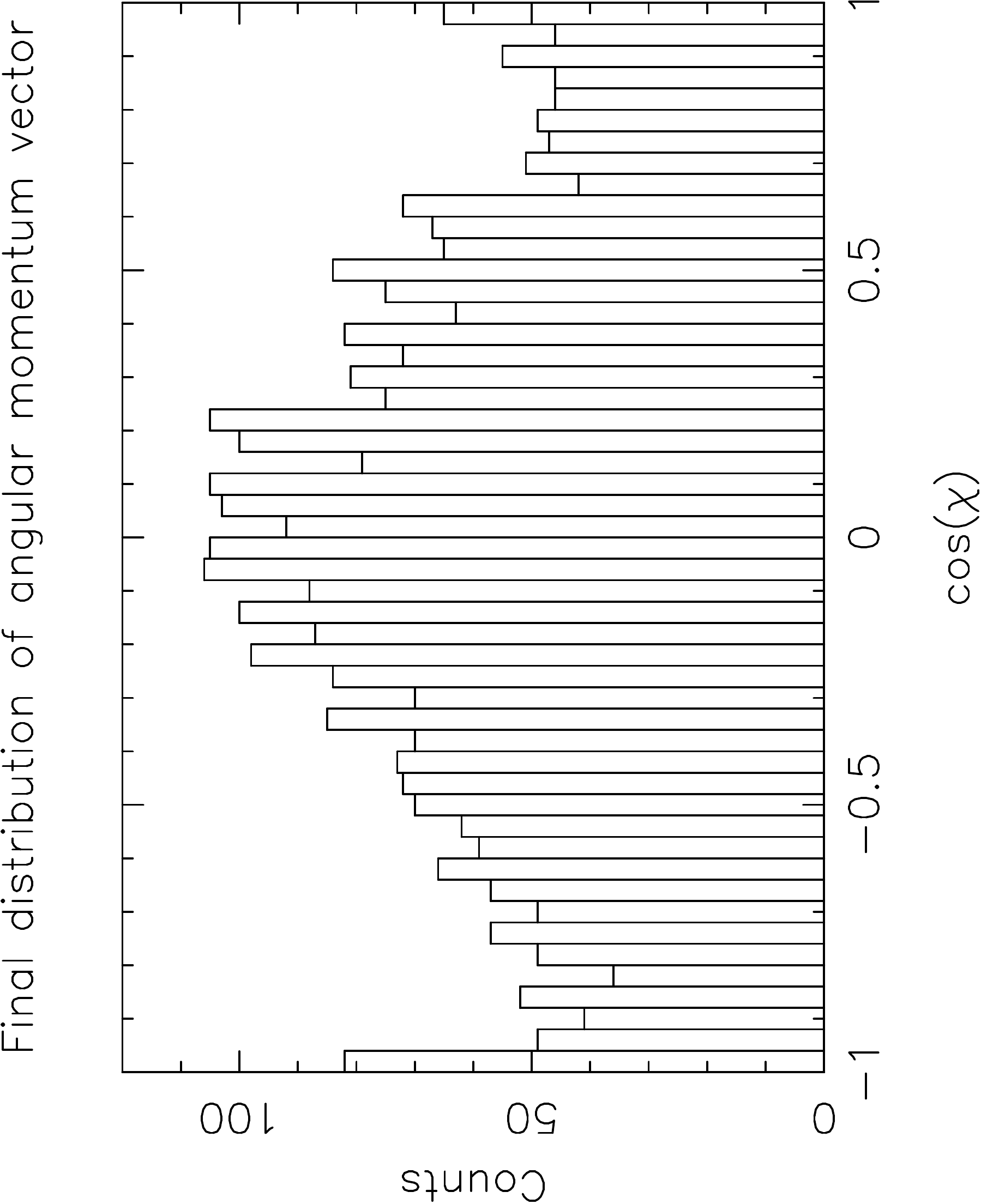}
\put (67,62) {\large PROLATE}
\end{overpic} &
\begin{overpic}[angle=-90,width=0.33\textwidth]{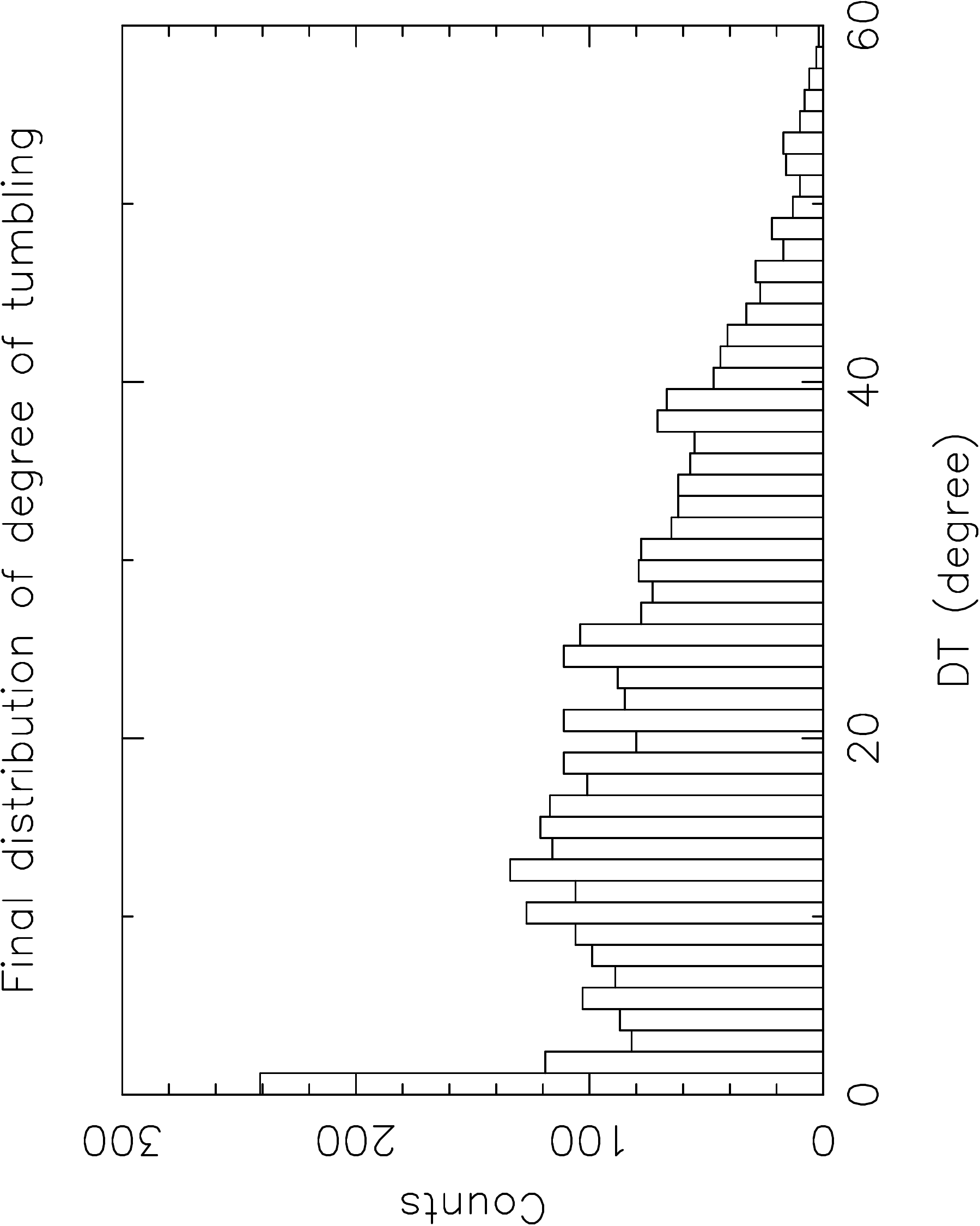}
\put (67,62) {\large PROLATE}
\end{overpic} &
\begin{overpic}[angle=-90,width=0.33\textwidth]{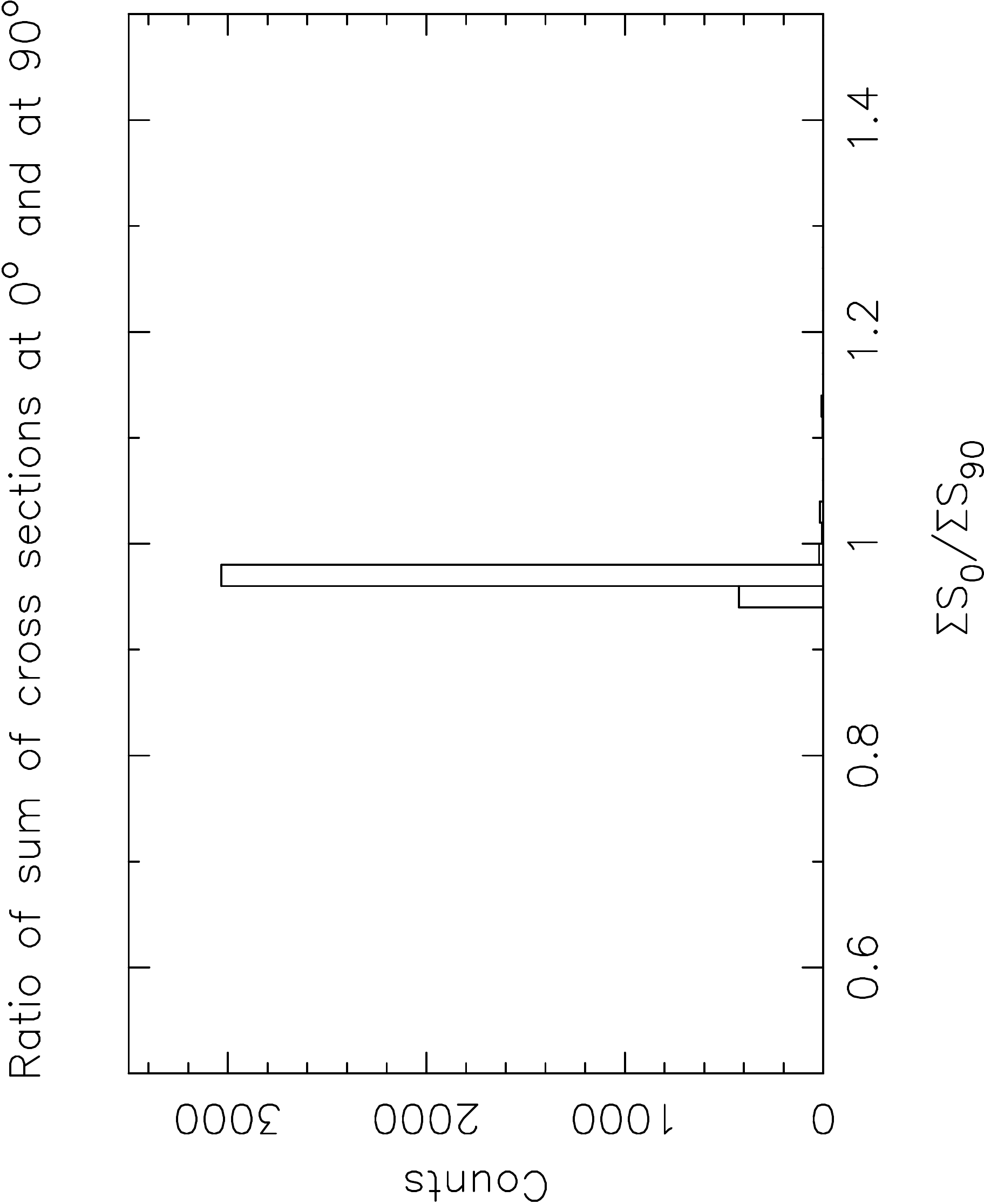}
\put (67,62) {\large PROLATE}
\end{overpic} \\
\end{tabular}
\caption{Results of dynamical parameters for the large sample of
  particles having $r_{e\!f\!f}$=1 cm at 67P perihelion. The upper row
  corresponds to oblate-shaped particles, and the lower row to
  prolate-shaped particles, as indicated. Angle $\chi$ is that formed
  by the final (at 50 km from the nucleus centre)
  angular momentum vector and the gas flux direction. A random
  distribution would display a constant count value for all
  $\cos(\chi)$. The degree of tumbling corresponds to the angle
  between the angular momentum vector and the spin axis of the particle at the end of
  integration. The rightmost panels display the histogram of the
  integrated projected area of the particle at 0$^\circ$ divided by
  that at 90$^\circ$. }
\label{figlargesample}
\end{figure*}

% Large sample table
\begin{table}
	\centering
	\caption{Dynamical parameters of the large sample of synthetic
          flattened and elongated particles of 
          $r_{e\!f\!f}$=1
          cm particles at perihelion ($r_h$=1.24 au).}
	\label{tablelargesample}
\begin{tabular}{lrr}
	  \hline
Parameter & Flattened & Elongated  \\
\hline
{\texttt{DT}} (deg) & 10$\pm$6 & 20$\pm$14 \\
$\mathbf{L}_{final}$ (deg) & 0$\pm$33 &
0$\pm$35 \\
$\nu_{f\!i\!n\!a\!l}$ (Hz) & 0.12$\pm$0.11 &
0.15$\pm$0.14  \\
$V_{f\!i\!n\!a\!l}$ (m s$^{-1}$) & 3.6$\pm$0.5 &
3.2$\pm$0.3 \\
$\Sigma S_0/\Sigma S_{90}$ & 1.00$\pm$0.03 &
0.97$\pm$0.02 \\  
\hline
\end{tabular}
\end{table}

The results for both sets of oblate- and prolate-shaped particles are
displayed in Figure~\ref{figlargesample}, and its corresponding numerical
results are shown in table~\ref{tablelargesample}. The results
obtained on final frequencies and velocities for flattened and
elongated shape model particles are in line with those
derived from the single shape {\it{Begonia}} and {\it{Glasenappia}}
models (see tables~\ref{size1cm} and~\ref{tablelargesample}). The final
direction of angular momentum vectors is expressed as the cosine
of the colatitude, $\chi$, of $\mathbf{L}_{final}$. In a random distribution of
directions, $\cos(\chi)$ would display a constant count value from -1
to 1. As with the single asteroidal shape models, the distribution of directions of
$\mathbf{L}_{final}$ is not random, the vectors mostly pointing to the perpendicular
to the gas flux direction. The degree of
tumbling, {\texttt{DT}}, are markedly different for both the oblate- and
prolate-shaped particles. In both case the distribution is bimodal,
with relative maxima near 0$^\circ$ and 15$^\circ$ in both cases,
although with different ratio between both maxima and distinct widths. For the flattened
particles, the histogram is similar to that computed by 
\cite[][see his figure 5]{2014A&A...568A..39C}. Regarding the
integrated area ratio ($\Sigma S_0/\Sigma S_{90}$) distribution, we 
confirm the value of this ratio obtained in the analysis
of the three separated model shapes as $\Sigma S_0/\Sigma S_{90}$
$\sim$1. This confirms that the observed phase function backscattering
effect observed is not caused by a mechanical alignment of the
particles, at least under the implicit model assumptions.

\section{Conclusions}

A model of the dynamical evolution of irregularly-shaped particles
in cometary comae environments has been developed. The model is based
on a quaternion-based scheme and includes the effect of both the
aerodynamic and radiative forces and torques. The radiative torques
were computed by a geometric optics approximation algorithm. For the
assumed absorbing particles the torques were found to agree with the
far more rigorous DDA calculations at sizes larger than 5 $\mu$m. The model has been 
validated with previous results on the dynamical behaviour of
spheroids subjected to aerodynamic forces and torques from
\cite{2017Icar..282..333I}. We calculated the  
dynamical parameters in the particular case of comet 67P coma, at 
perihelion ($r_h$=1.24 au) and at far heliocentric distances ($r_h$=2
au), for two particle sizes,
1 mm, and 1 cm, and three asteroid-like shape model particles (rounded, elongated,
and flattened) up to a distance of 50
km from the nucleus. The
effect of the radiative torques is completely negligible up to 50 km
from the nucleus, but we cannot exclude some effect at farther
distances. Overall, the final rotational frequencies obtained for such particles vary
between 0.02 Hz and 2.2 Hz, being a function of size, shape, and
heliocentric distance. These values are in line with the measured
frequencies in the 67P coma by OSIRIS images, which were found to the
caused precisely by particles in the mm to cm range \citep{2021MNRAS.504.4687F}.  For effective
radii $\lesssim$10 $\mu$m, the 
rotation frequencies exceed 500 Hz, and the time step needed to
guarantee stability in the solution must be so small that
long-term integrations are precluded in practice. 

Based on those results of the three shape model particles, and in
order to give a more robust estimate of the derived dynamical
parameters, we
conducted a more complete statistics on the dynamical behaviour of
irregularly shaped particles by building up a large database of oblate-like and
prolate-like shape model particles distributed in wide axes ratio
distributions.  The results derived from such analysis are consistent
with those performed on the smaller satistical sample. The 
dynamical evolution of irregularly-shaped particles is different to
that found by symmetrical particles in the far more complex
rotational motion that characterises such irregular particles. The
evolution is chaotic during the first few kilometres from the nucleus
surface, and eventually becomes 
more regular. The direction of the angular momentum vector at the end
of the integration time is close to the perpendicular to the gas flow,
and the degree of tumbling (angle between the angular momentum vector
and spin axis) vary between $\sim$10$^\circ$ and
$\sim$20$^\circ$ in average, for flattened and elongated particles,
respectively, independently of size and heliocentric distance. These results
agree very well with the analysis of the dynamics of meteoroid rotation
performed by \cite{2014A&A...568A..39C}. On the other hand, the
terminal velocities (at a distance 50 km of the nucleus) do not depend
strongly on the aspect ratio, being similar among
different shape model particles.  

In order to shed some light on the possible implications of the
particle dynamical behaviour on the measured comet dust phase function
and its backscattering enhancement, an analysis of the computed ratio
of the sum of the particle projected areas in the
sun-comet line direction (maximum gas flux) and in its perpendicular direction
($\Sigma S_0/\Sigma S_{90}$) along the trajectory has been made. This ratio
approaches unity for all model particles considered, so that the shape
of the phase function would not be altered as a consequence of a
geometrical effect linked to particle alignment, at least for the model
assumptions adopted.

\section*{Acknowledgements}

We thank the anonymous referee for the careful review of
    the manuscript.

We are indebted to P. Cignoni and collaborators for the use of 
their MeshLab software, and to B. Draine and P. Flatau for making
available their Discrete Dipole Approximation DDSCAT code. 

FM, DG, and OM acknowledge financial support from the State Agency for
Research of 
the Spanish MCIU through the "Center of Excellence Severo Ochoa" award
to the Instituto de Astrof\'\i sica de Andaluc\'\i a (SEV-2017-0709).
FM and DG also acknowledge financial support from the  Spanish  Plan
Nacional  de  Astronom\'\i a  y  Astrof\'\i sica  LEONIDAS
project RTI2018-095330-B-100,  and  project  P18-RT-1854  from  Junta
de  Andaluc\'\i a.

\section{Data availability}
This work uses simulated data, generated as detailed in the text.

\bibliographystyle{mnras}
\bibliography{paper-final-arXiv}

% Don't change these lines
\bsp	% typesetting comment
\label{lastpage}
\end{document}